\documentclass{aa}

\usepackage{graphicx}
\usepackage{txfonts}
\usepackage{color}

\begin{document}

\title{Lopsided galactic bars}

\author{Ewa L. {\L}okas
}

\institute{Nicolaus Copernicus Astronomical Center, Polish Academy of Sciences,
Bartycka 18, 00-716 Warsaw, Poland\\
\email{lokas@camk.edu.pl}}


\abstract{
Most of the observed and simulated galactic bars are symmetric in the face-on view. However, there are indeed cases of
bars that are off-center with respect to the disk or have an asymmetric shape. The only well-known example showing both
these features is the Large Magellanic Cloud. We report on the identification of several lopsided galactic bars in the
Illustris TNG100 simulation found among a sample of elongated bar-like galaxies studied in the past. The bars show a
clear asymmetry in the face-on view, which is in the shape of a footprint. We measured the evolution of the different
parameters of the bars' shape and asymmetry as a function of time and find that the asymmetry is preserved for a few
Gyr. It can grow together with a bar or appear later, after bar formation. We considered two scenarios leading to the
formation of lopsided bars using controlled simulations. In the first, a Milky Way-like galaxy interacts with a massive
companion placed on a radial orbit in the plane of the disk and perpendicular to the orientation of the bar at the time
of the first passage. In the second, the galaxy initially has an off-center disk and the growth of the bar and its
asymmetry is more similar to the one found in IllustrisTNG galaxies, where it is also preceded by the presence of an
asymmetric disk. It is thus possible that lopsided bars are born in lopsided disks, although in some cases, the time
difference between the occurrence of the asymmetry in the two components is quite large.}


\maketitle

\section{Introduction}

A significant fraction of galaxies in the Universe appear to be barred \citep{Buta2015}. This fraction depends
on the sample selection, stellar mass, gas content, environment, and redshift and it may reach as high as 60\%
\citep{Sheth2008, Skibba2012, Melvin2014, Diaz2016}. The bars are typically symmetric, elongated structures embedded
in well-formed disks. Similar structures are reproduced in simulations of galaxy formation, independently of their
origin, whether they are formed via bar instability in isolation \citep{Hohl1971, Ostriker1973, Sellwood1981,
Athanassoula2002, Debattista2006, Athanassoula2013} or as a result of tidal interactions with other objects
\citep{Gerin1990, Noguchi1987, Miwa1998, Lokas2014, Lokas2018}.

The symmetry seems to be an inherent feature of the bars, with the notable exception of the bar in the Large Magellanic
Cloud (LMC). While this bar has typically been considered lopsided because of its off-center position with respect to the
disk, it seems to be asymmetric also in its intrinsic shape, as demonstrated in the studies of \citet{Marel2001} and
\citet{Jacyszyn2016}. A whole class of Magellanic-type galaxies was later identified \citep{Odewahn1994}, but this name
again has referred to their bars being offset with respect to the disks, rather than being asymmetric in shape. Recent
simulations of the LMC formation have also been aimed at reproducing the off-center bar, rather than the asymmetric one
\citep{Bekki2009, Besla2012, Pardy2016}. The phenomenon is not rare; \citet{Kruk2017} identified as many as 270
late-type galaxies with off-center bars using Sloan Digital Sky Survey (SDSS) images and Galaxy Zoo morphologies.

The studies of asymmetric structures in galaxies have thus far been focused on lopsided disks rather than bars since there is much
more observational material on hand in the former case \citep{Jog2009}. For example, \citet{Rix1995} found that out of 18
face-on spirals they studied, about 1/3 were substantially lopsided with significant values of odd Fourier modes. These
results were later confirmed and extended to 147 galaxies of the OSUBGS sample by \citet{Bournaud2005}. More recently,
\citet{Zaritsky2013} confirmed the high incidence of lopsidedness in their sample of 167 galaxies, spanning a wide range
of luminosities and morphologies. \citet{Bournaud2005}, \citet{Mapelli2008} and \citet{Ghosh2021} considered the
possible origin of lopsidedness in galactic disks using simulations of galaxy evolution. The conclusion from these
studies was that although interactions such as flybys or minor mergers are a plausible cause for such distortions,
they cannot explain all occurrences of this phenomenon given its presence in isolated galaxies as well.

\begin{figure*}
\centering
\includegraphics[width=5.7cm]{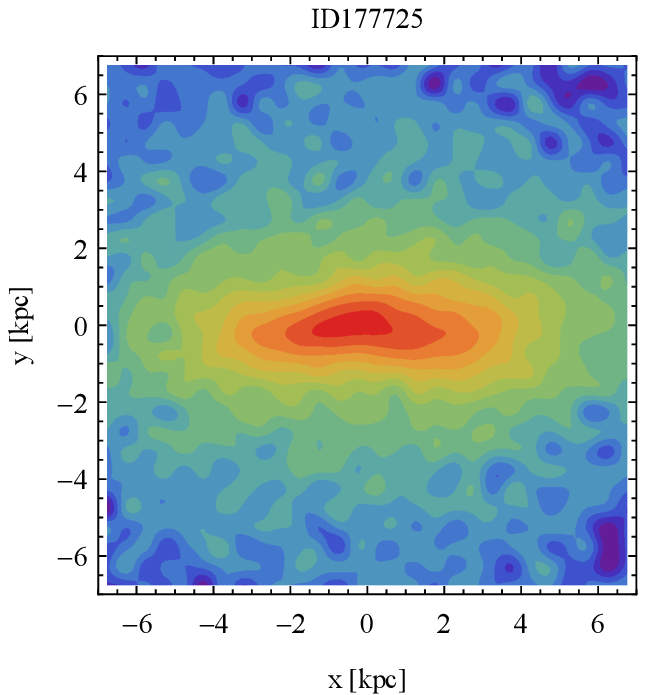}
\includegraphics[width=5.7cm]{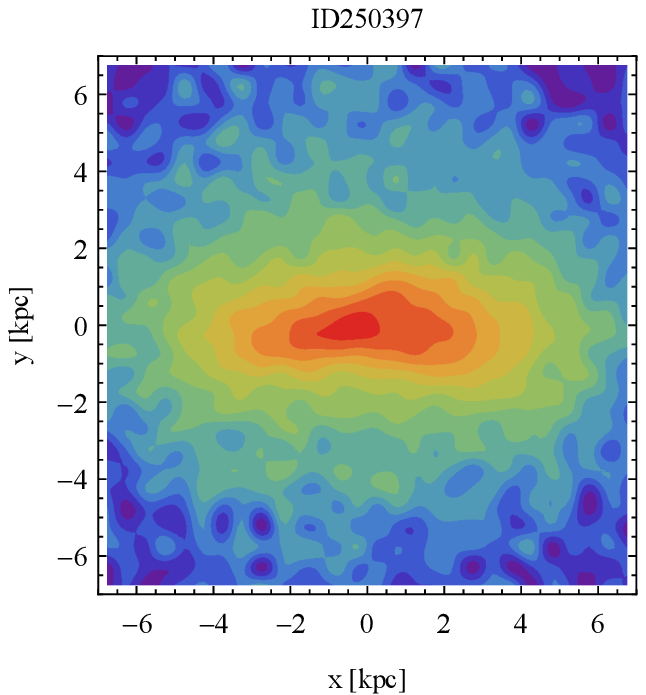}
\includegraphics[width=5.7cm]{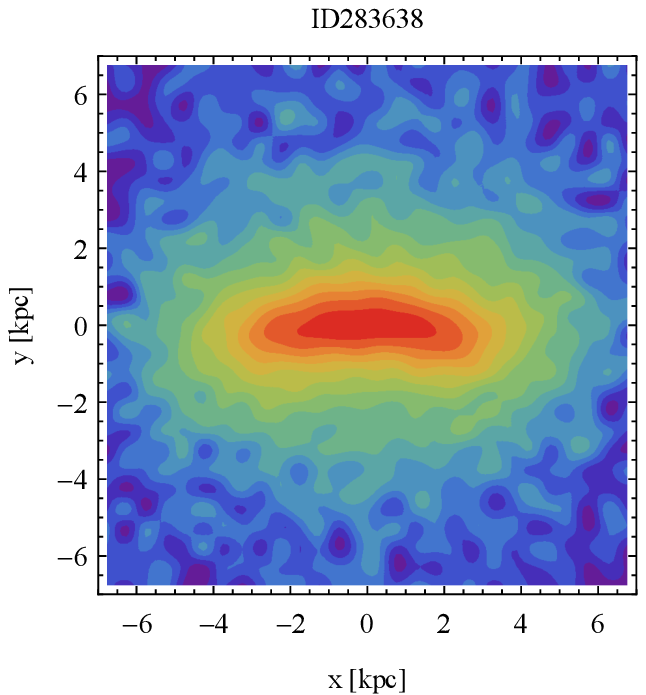} \\
\vspace{0.3cm}
\includegraphics[width=5.7cm]{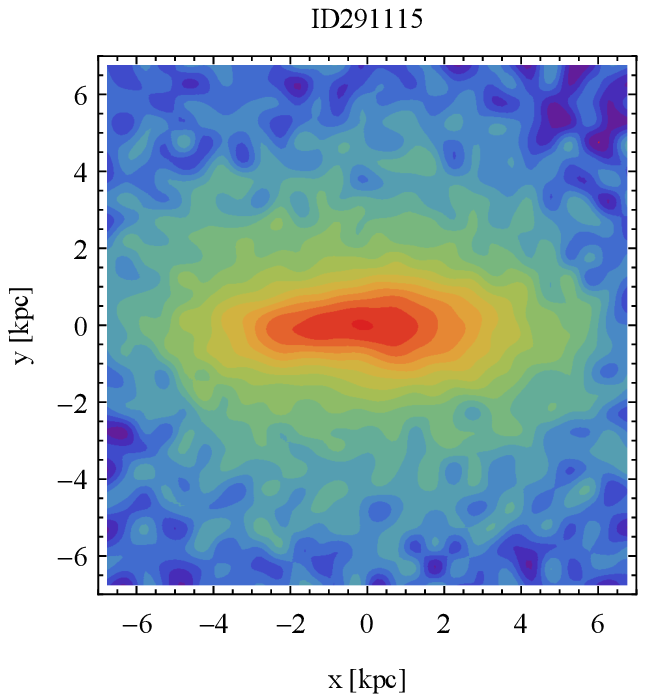}
\includegraphics[width=5.7cm]{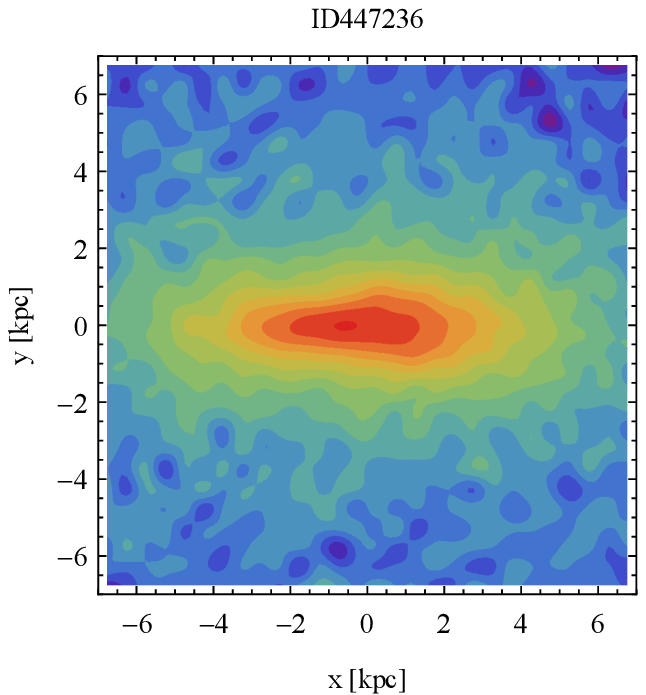}
\includegraphics[width=5.7cm]{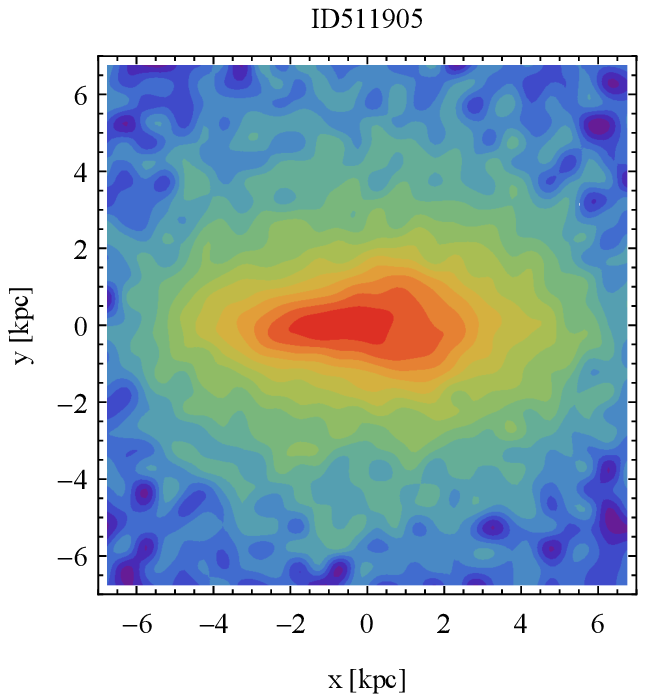}
\caption{Surface density distribution of the stellar components of the lopsided bars from IllustrisTNG, in the face-on
view at the present time. The surface density, $\Sigma,$ is normalized to the central maximum value in each case and
the contours are equally spaced in $\log \Sigma$.}
\label{surden}
\end{figure*}

\begin{table*}
\caption{Properties of lopsided bars from IllustrisTNG at $z=0$.}
\label{properties}
\centering
\begin{tabular}{c c r r c c c c c c c c}
\hline\hline
ID \     &  $M_*$                &  $M_{\rm g}$ \ \ \ \  & $M_{\rm dm}$ \ \ \ \ &$r_{1/2}$& $b/a$  & $c/a$ & $T$  & $A_2$ & $A_1$& $A_3$& $A_5$ \\
         & [$10^{10}$ M$_\odot$] & [$10^{10}$ M$_\odot$] & [$10^{11}$ M$_\odot$]& [kpc]   &        &       &      &       &      &      &       \\ \hline
177725   &  3.43                 &  0.00 \ \ \ \         &  2.62 \ \ \ \        & 2.75    &  0.55  &  0.41 & 0.84 & 0.44  & 0.04 & 0.10 & 0.08  \\
250397   &  2.32                 &  0.00 \ \ \ \         &  1.29 \ \ \ \        & 2.52    &  0.58  &  0.39 & 0.79 & 0.40  & 0.03 & 0.08 & 0.08  \\
283638   &  3.23                 &  1.10 \ \ \ \         &  4.18 \ \ \ \        & 2.08    &  0.50  &  0.38 & 0.88 & 0.48  & 0.02 & 0.07 & 0.08  \\
291115   &  3.62                 &  4.03 \ \ \ \         &  6.38 \ \ \ \        & 2.34    &  0.51  &  0.41 & 0.89 & 0.45  & 0.03 & 0.09 & 0.08  \\
447236   &  4.83                 & 14.22 \ \ \ \         & 17.97 \ \ \ \        & 3.23    &  0.59  &  0.39 & 0.76 & 0.44  & 0.01 & 0.07 & 0.07  \\
511905   &  3.59                 &  1.39 \ \ \ \         & 11.90 \ \ \ \        & 2.09    &  0.58  &  0.42 & 0.80 & 0.36  & 0.02 & 0.11 & 0.09  \\
\hline
\end{tabular}
\end{table*}

In recent years, studies of the origin of galaxy morphology have entered a new era with the advent of cosmological
simulations that produce large samples of galaxies with a resolution that makes it possible to discern their basic morphological
features. One such set of simulations of galaxy formation is that of the IllustrisTNG collaboration
\citep{Springel2018, Marinacci2018, Naiman2018, Nelson2018, Pillepich2018}. Various studies performed thus far have
demonstrated that these simulations are able to reproduce many of the observed properties of galaxies, including their
morphologies \citep{Nelson2018, Genel2018, Rodriguez2019}. IllustrisTNG (and the earlier Illustris) simulations have
been used, in particular, to study the formation and properties of galactic bars \citep{Peschken2019, Rosas2020,
Zhou2020, Zhao2020}. Recently, we extended these efforts \citep{Lokas2021} to include an additional class of
bar-like galaxies selected on the basis of a single criterion: namely, that the stellar component is sufficiently elongated.
The 277 galaxies chosen in this way differ from the typical barred galaxies studied earlier in the sense that these
are not bars embedded in disks but, instead, almost their whole stellar component is elongated and prolate.

The detailed properties of these objects and their evolutionary histories are presented in \citet{Lokas2021}
with three fiducial examples. During the course of this study, we also inspected the face-on mock images of these
objects and discovered among them ten cases of asymmetric or lopsided bars, which comprise the subject of the present paper.
We further restricted the sample considered here to six objects by rejecting three for which the degree of
asymmetry was rather low and one that is strongly distorted due to an ongoing merger. Although some of the bars
considered in the earlier studies based on IllustrisTNG \citep{Rosas2020, Zhao2020} also show some degree of asymmetry,
these bars are generally weaker and their asymmetry is also less pronounced. Therefore, we restricted the present study to bar-like objects that are significantly stronger bars overall. In the next section, we describe the properties
of the lopsided bars identified among the bar-like galaxies from IllustrisTNG. In Section~3, we present
controlled simulations of possible scenarios for the formation of lopsided bars. Our discussion is presented in the
last section.

\begin{figure*}
\centering
\includegraphics[width=18cm]{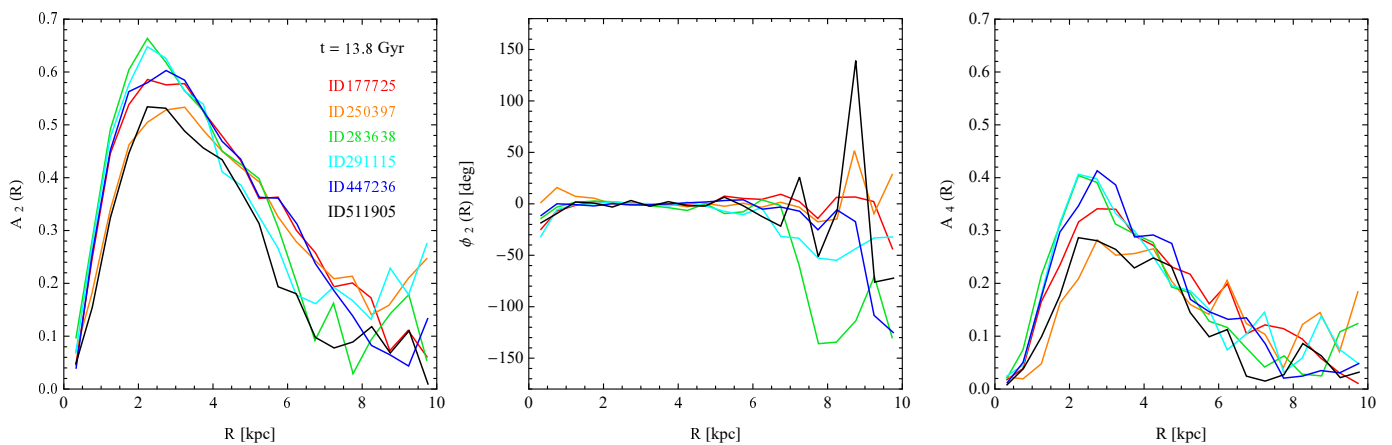}
\caption{Profiles of bar mode $A_2 (R)$ (left panel), the phase of the bar mode $\phi_2 (R)$ (middle panel),
and the $m = 4$ mode $A_4 (R)$ (right panel) for IllustrisTNG galaxies with lopsided bars at the present time.
Measurements were carried out in bins of $\Delta R = 0.5$ kpc.}
\label{a2a4phaseprofiles}
\end{figure*}

\section{Lopsided bars in IllustrisTNG}

In this study, we used the publicly available simulation data from the 100 Mpc box of the IllustrisTNG project, as
described by \citet{Nelson2019}. This simulation has at redshift $z = 0$ the Plummer-equivalent gravitational softening
length for the collisionless components (dark matter and stars), $\epsilon = 0.74$ kpc, and the median stellar particle
mass, $1.6 \times 10^6$ M$_\odot$ \citep{Pillepich2018}. The TNG100 run provides a sufficiently numerous sample of
galaxies with a good enough resolution for analyzing different morphological samples. We used the subsample of bar-like
galaxies from \citet{Lokas2021} selected at $z=0$ that had total stellar masses greater than $10^{10}$ M$_\odot$
($10^4$ stellar particles per object) and that obeyed the morphological condition of the intermediate to longest axis ratio
$b/a$ of the stellar component lower than 0.6. The axis ratios were estimated from the eigenvalues of the mass tensor
of the stellar mass within two stellar half-mass radii, $2 r_{1/2}$, as described in \citet{Genel2015}.

Among the sample of 277 bar-like galaxies, we found (via visual inspection) six objects that showed a significant asymmetry
of the bar in the face-on image and were otherwise regular. The surface density distributions of the stars for these
six galaxies at the present time ($z = 0$, the last simulation output) are shown in Fig.~\ref{surden}. In each case,
the images are rotated so that the longest axis of the stellar component and the bar are along the $x$ axis of the plot
and centered on the dynamical center of the galaxy, which for Illustris galaxies is determined as the position of the
particle with the minimum gravitational potential energy. The asymmetry is clearly visible in these images in the form
of a footprint-like shape (or banana-like, as in one case).

Except for the asymmetry of their bars, the galaxies are typical examples of the bar-like galaxies discussed in
\citet{Lokas2021} in terms of their masses, shapes, and evolutionary histories. Table~\ref{properties} lists the most
important properties of the galaxies. The first column of the Table gives the identification number of the galaxy
according to the subhalo catalogs of the Illustris TNG100 simulation. The next three columns list the stellar, gas, and
dark matter total masses. The dark and stellar masses of the galaxies are similar to those of the Milky Way (MW) or
slightly smaller. The majority of the galaxies also contain a significant amount of gas, but mostly in the outskirts.

The evolutionary histories of the six galaxies with lopsided bars are quite different, as was the case for the general
population of bar-like objects discussed previously by \citet{Lokas2021}. The first two galaxies in the Table are members of
clusters and, therefore, they already lost all their gas and most of their dark matter due to ram pressure and tidal stripping. The
second pair of galaxies shown in the table also interacted with more massive structures, but much more weakly, thus retaining
some of the gas and most of the dark matter. The last pair evolved mostly in isolation from more massive objects growing in
mass until now.

In order to describe the shape of the bars quantitatively, it is useful to calculate different modes of the
Fourier decomposition of the surface density distribution of stellar particles projected along the short axis, $A_m (R)
= | \Sigma_j m_j \exp(i m \theta_j) |/\Sigma_j m_j$, where $\theta_j$ is the azimuthal angle of the $j$th star, $m_j$
is its mass, and the sum goes up to the number of particles in a given radial bin.
In the left panel of Fig.~\ref{a2a4phaseprofiles}, we plot the dependence of the
bar mode, $A_2,$ on the cylindrical radius at the present time. All the profiles show a behavior characteristic of bars,
in spite of their asymmetry, that is, the values of $A_2$ first increase with radius, reaching a maximum and then
decreasing. Because of the asymmetry, one side of the bar contributes more to the $A_2$ value, but the
$A_2$ mode is still the dominant one and departures from the symmetry are to be quantified with the odd modes, as discussed
below. The bars are quite strong, with the maximum $A_2$ values above 0.5 in all cases.

\begin{figure*}
\centering
\includegraphics[width=18cm]{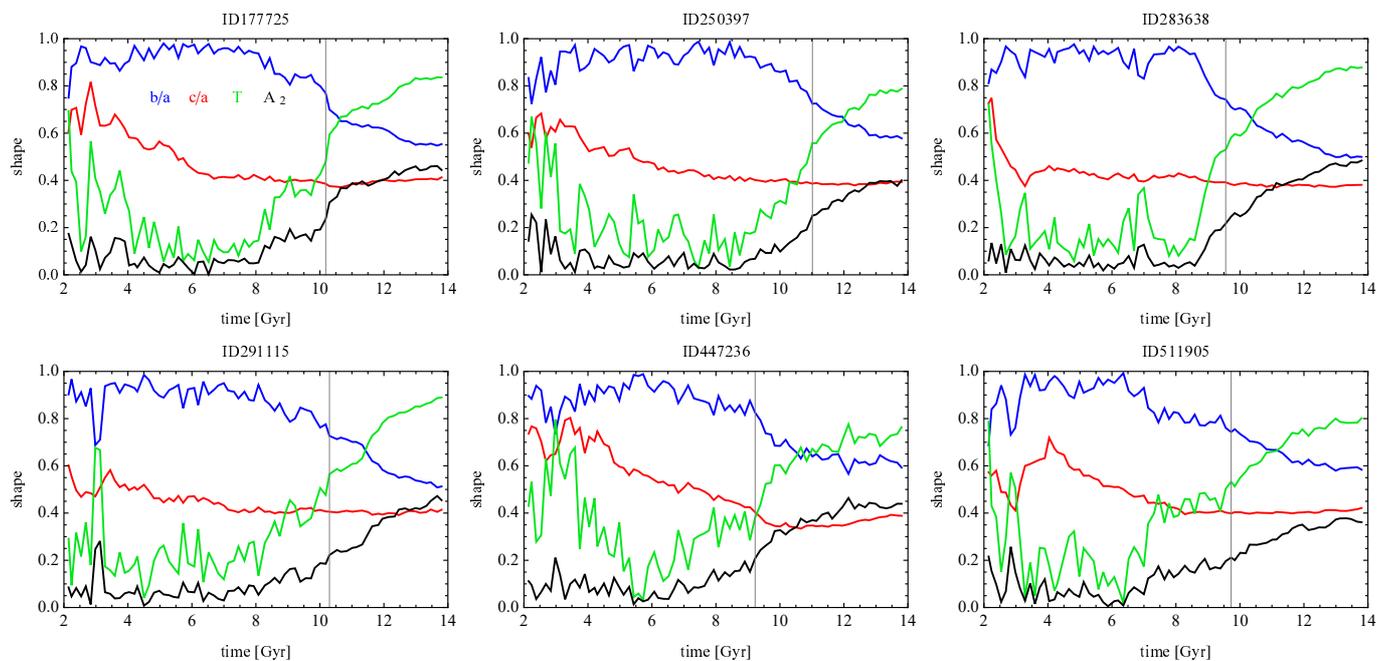}
\caption{Evolution of the shape of the lopsided bars from IllustrisTNG. Blue and red lines plot the intermediate to
longest $b/a$ and the shortest to longest $c/a$ axis ratios, respectively. Green and black lines show the evolution
of the triaxiality parameter, $T,$ and the bar mode, $A_2$. Vertical gray lines indicate the times of bar formation.}
\label{shape}
\end{figure*}

The radii where the $A_2$ profile drops to half the maximum can be used to estimate the length of the bar and we see
that these lengths are on the order of 6 kpc for all bars at the present time, in agreement with the visual
impression from Fig.~\ref{surden}. The middle panel of Fig.~\ref{a2a4phaseprofiles} shows the profiles of the
phase angle of the $m = 2$ Fourier mode, $\phi_2 (R)$. We see that its values are constant and close to zero out to
radii of the order of 6-7 kpc. Since such radii can also be used as estimates of the bar length we find them to agree
with the lengths estimated from the behavior of the $A_2$ profile. Finally, the right panel of
Fig.~\ref{a2a4phaseprofiles} shows the profiles of the $m = 4$ Fourier modes $A_4$ known to be typically excited in
bars, in addition to $m = 2$ modes. Their values are significant, although smaller than $A_2$, with maximum values above
0.2 for all galaxies.

Figure~\ref{shape} illustrates the evolution of the different measures of the shape of the galaxies. As mentioned
before, the axis ratios and all the other properties discussed below were estimated using stars contained within two
stellar half-mass radii, $2 r_{1/2}$, listed in the fifth column of Table~\ref{properties} for the present time. The
six panels of the figure plot the evolution of the intermediate to longest $b/a$ and shortest to longest $c/a$ axis
ratio as a function of time as well as the triaxiality parameter, $T = [1-(b/a)^2]/[1-(c/a)^2]$, for the six lopsided
bars. The final values of $b/a$, $c/a,$ and $T$ for all galaxies are given in columns 6-8 of Table~\ref{properties}. In
all cases, the ratio $b/a$ is initially close to unity and $T$ is close to zero, so the galaxies are oblate disks. At
some point, however, the $b/a$ start to decrease and $T$ to increase signifying a transition from an oblate to a
prolate shape, which indicates the formation of a bar.

The strength of the bar in the form of the $m=2$ mode for stars within two stellar half-mass radii, $2 r_{1/2}$, is
also plotted in Fig.~\ref{shape} and we can see that its evolution traces that of the triaxiality. The final values of
$A_2$ are listed in column 9 of Table~\ref{properties}. In vertical gray lines, we marked the times where $A_2$
crosses the value of 0.2 for the first time, which can be adopted as the time of bar formation. This choice,
while it may appear rather arbitrary, reflects the presence of an elongated shape and is useful as an operational definition for
the onset of the bar.

\begin{figure*}
\centering
\includegraphics[width=18cm]{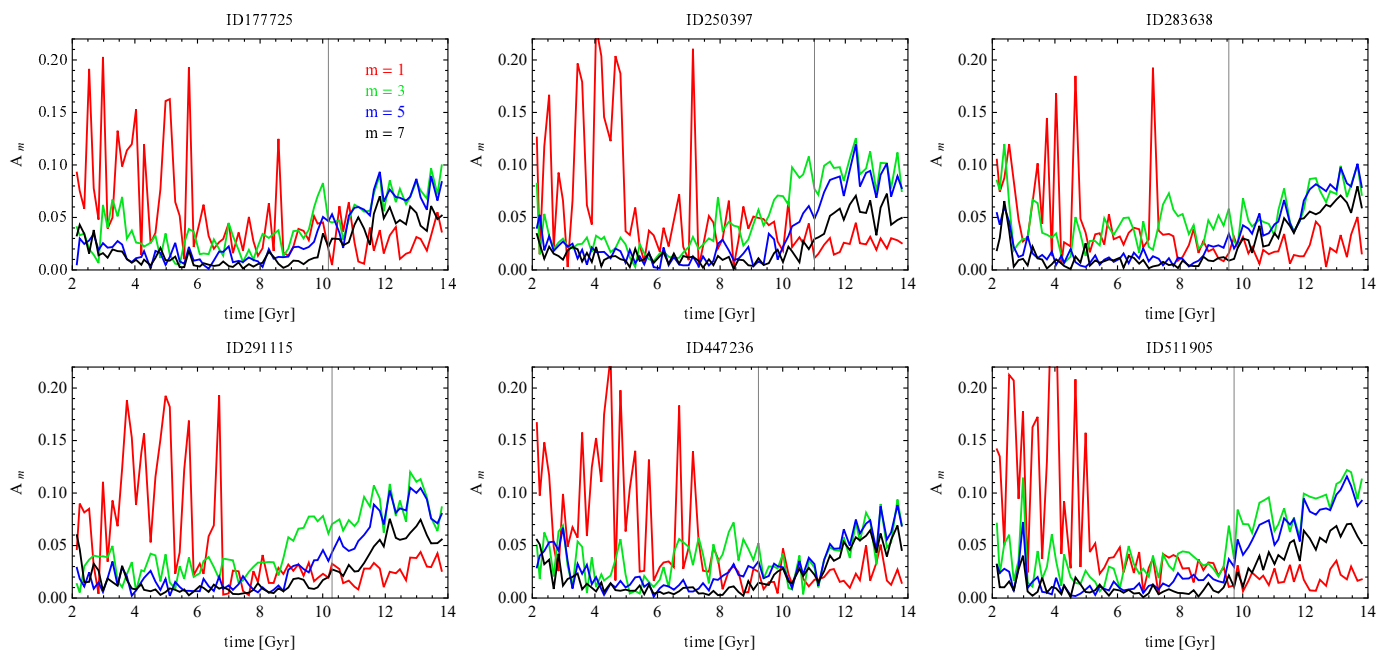}
\caption{Odd Fourier modes of the stellar component of IllustrisTNG galaxies measured within two stellar
half-mass radii, $2 r_{1/2}$, as function of time. The red, green, blue, and black lines show the results for $m = 1$,
3, 5, and 7, respectively. Vertical gray lines indicate the times of bar formation.}
\label{oddmodestime}
\end{figure*}

\begin{figure}[ht!]
\centering
\hspace{0.4cm}
\includegraphics[width=3.5cm]{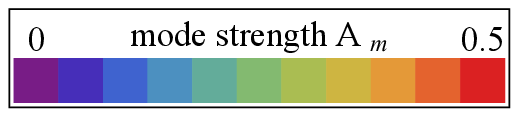}
\vspace{0.1cm}
\includegraphics[width=8.9cm]{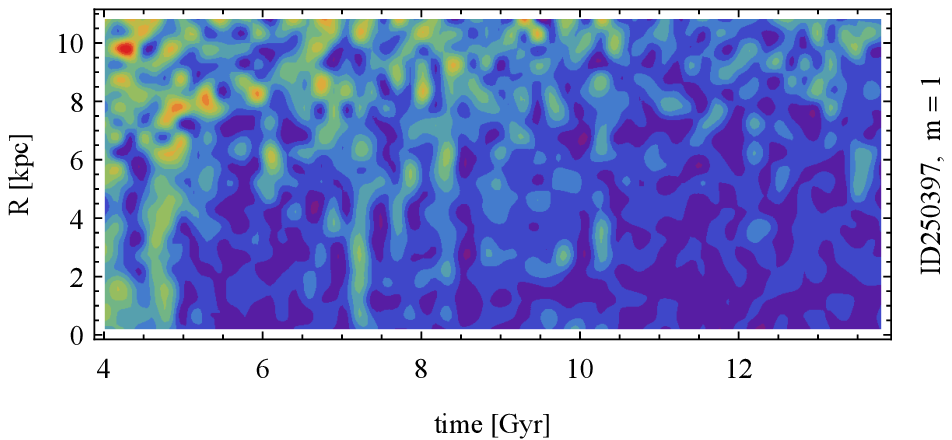}
\vspace{0.1cm}
\includegraphics[width=8.9cm]{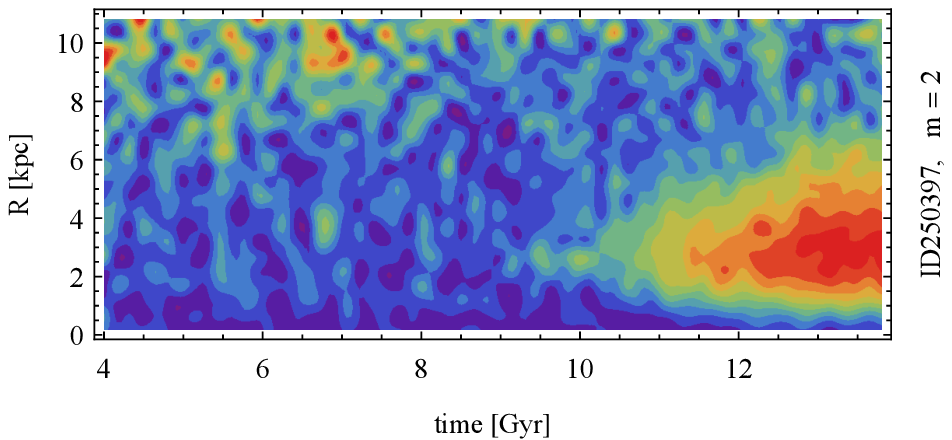}
\vspace{0.1cm}
\includegraphics[width=8.9cm]{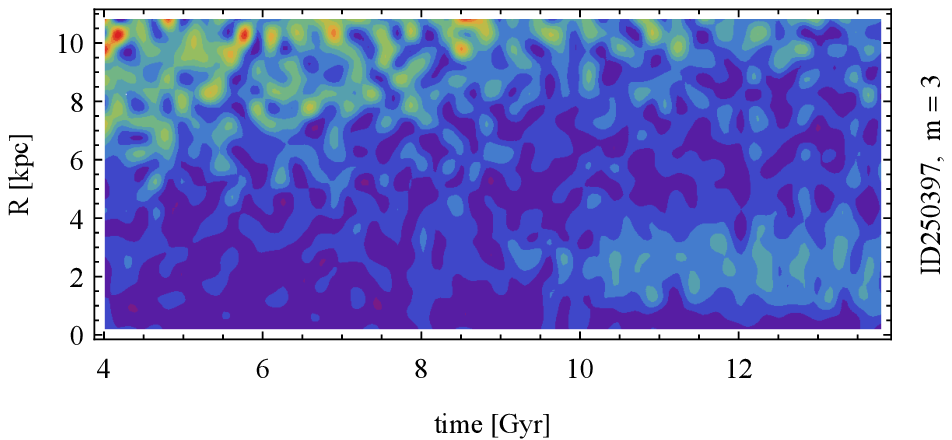}
\vspace{0.1cm}
\includegraphics[width=8.9cm]{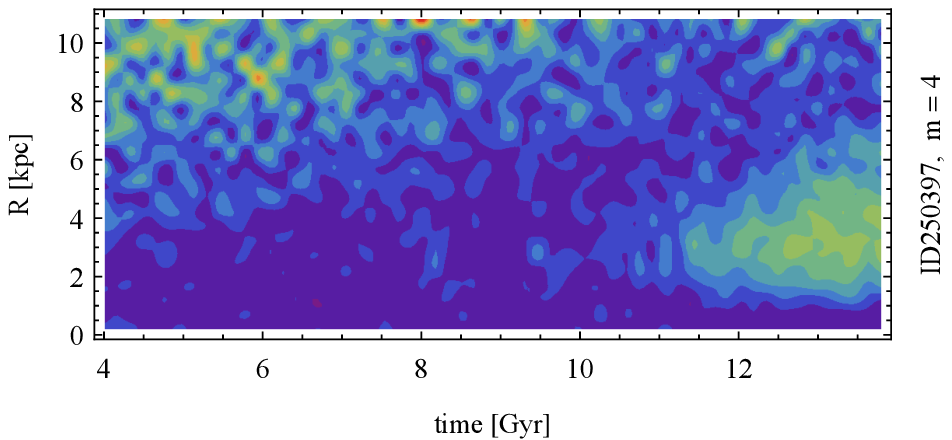}
\vspace{0.1cm}
\includegraphics[width=8.9cm]{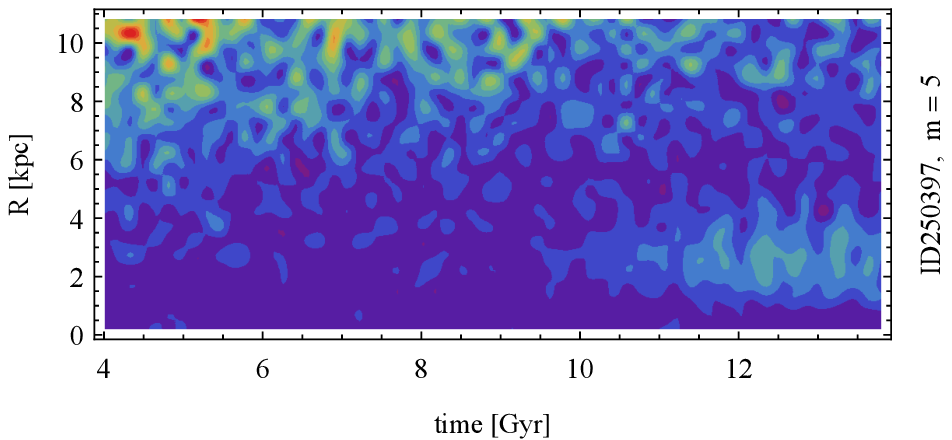}
\caption{Evolution of the profiles $A_m (R)$ over time for galaxy ID250397 from IllustrisTNG. The five panels
from top to bottom show the results for $m = 1$ to 5.}
\label{a2modestime1}
\end{figure}

\begin{figure}[ht!]
\centering
\hspace{0.4cm}
\includegraphics[width=3.5cm]{legend24.eps}
\vspace{0.1cm}
\includegraphics[width=8.9cm]{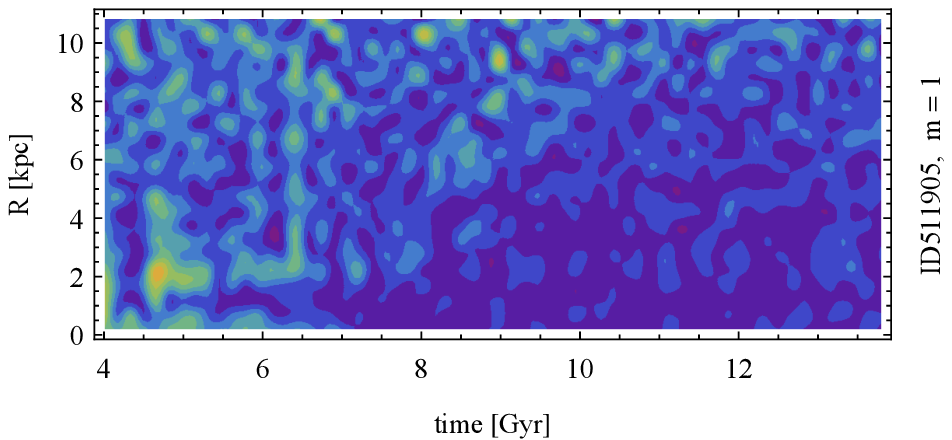}
\vspace{0.1cm}
\includegraphics[width=8.9cm]{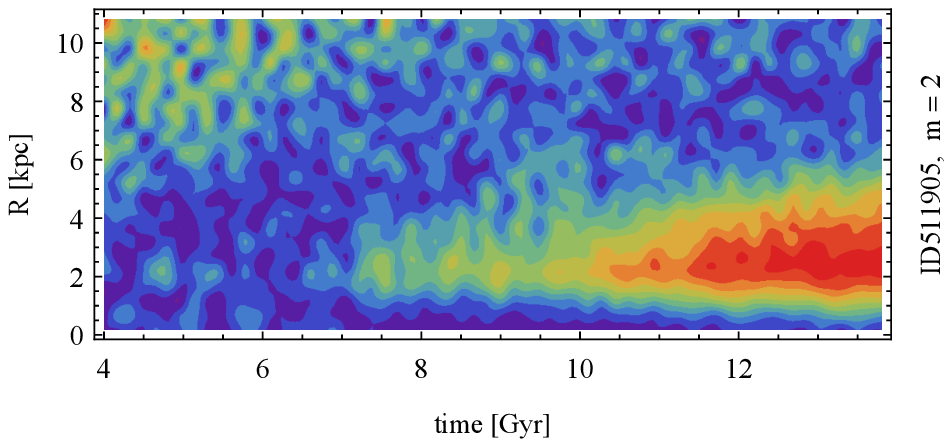}
\vspace{0.1cm}
\includegraphics[width=8.9cm]{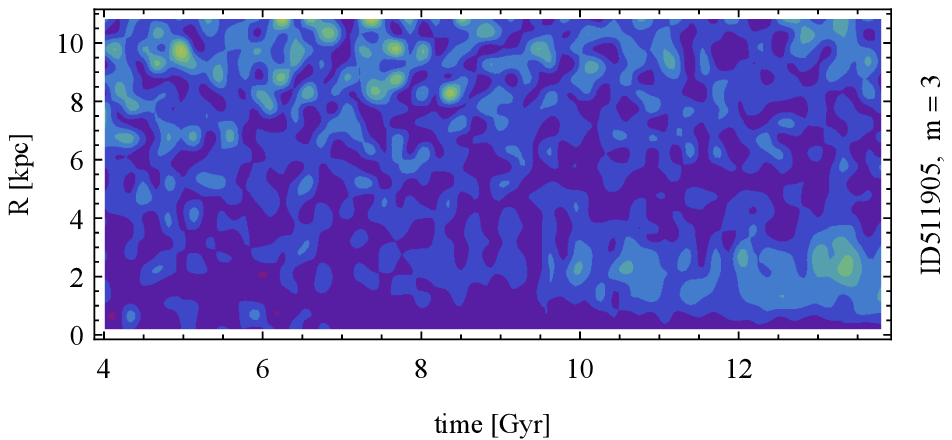}
\vspace{0.1cm}
\includegraphics[width=8.9cm]{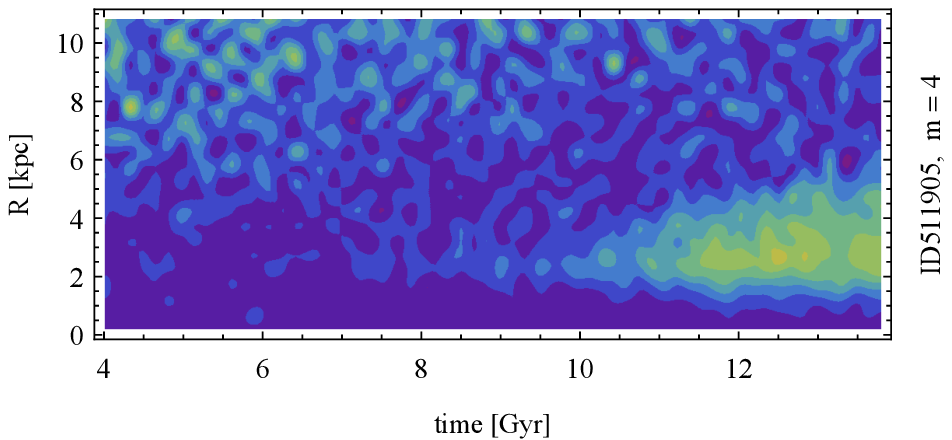}
\vspace{0.1cm}
\includegraphics[width=8.9cm]{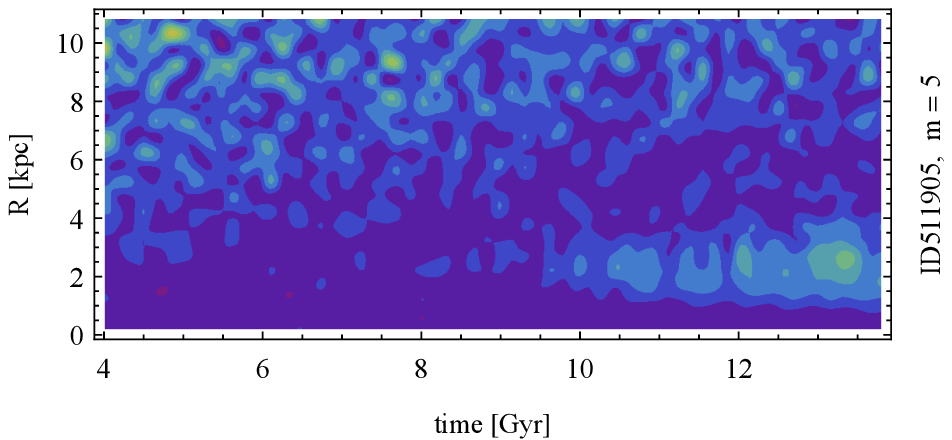}
\caption{Evolution of the profiles $A_m (R)$ over time for galaxy ID511905 from IllustrisTNG. The five panels
from top to bottom show the results for $m = 1$ to 5.}
\label{a2modestime2}
\end{figure}

The asymmetry of the distribution of stars in the bars can be described by the odd modes of the Fourier decomposition.
Figure~\ref{oddmodestime} shows the evolution of $A_m$ for $m = 1$, 3, 5, and 7 for the six galaxies in the
same time range as in Fig.~\ref{shape}. Columns 10-12 of Table~\ref{properties} list the final values of $A_1$, $A_3$
and $A_5$. We can see that the $m=1$ mode, signifying asymmetry in the disks, varies strongly with numerous
peaks at early stages of the evolution when the galaxies are forming and experience mergers and accretion events, but is
usually subdominant at later stages when the bar is already in place. Then $A_3$ and $A_5$ take largest values and
$A_7$ is again smaller. Thus, $A_3$ and $A_5$ can be considered as the best signatures of the lopsidedness of the bars and
we note that they preserve the values of the order of 0.1 for a few Gyr. This means that the lopsidedness is not a
transient feature but, rather, a stable property of the bars. We also note that the asymmetry can occur at different stages of
bar formation. In some cases, the asymmetry develops after the bar is formed (ID447236) and sometimes around the time
of bar formation (ID250397), which means that the bar is born asymmetric.

We have found that the hierarchy of the odd $A_m$ modes is different from the one found in lopsided disks where
the $m = 1$ mode dominates \citep{Rix1995, Bournaud2005}. The reason for this is that in the later stages of
the evolution of the galaxies studied here the disk is not lopsided and the dominant structural property is the
presence of the bar, as confirmed by the highest values of the $m = 2$ and $m = 4$ modes. The $m = 3$ and $m = 5$ modes
can be understood as corrections to the even bar modes describing the departures of the bar from symmetry.
Geometrically, non-zero $m = 3$ and $m = 5$ modes can be associated with the triangular and pentagonal shape of the
bar, respectively.

In order to fully explore the information contained in the Fourier modes, in
Figs.~\ref{a2modestime1}-\ref{a2modestime2}, we show their dependence on both time and radius for two galaxies with
the largest asymmetry that is also preserved for a long time, namely, ID250397 and ID511905. The figures plot the
evolution of the profiles of the Fourier modes, $A_m,$ in time for $m = 1$ to $m = 5$ (from top to bottom) in a
color-coded format. The bin size in radius is $\Delta R = 0.5$ kpc (as before) and the average time difference between
IllustrisTNG snapshots is $\Delta t = 0.16$ Gyr. The second and fourth panels of these figures illustrate the formation
of the bar in terms of $A_2 (R, t)$ and $A_4(R, t)$. They can also be used to estimate the bar length at different
times by finding a radius where the $A_2$ value drops to half its maximum.

For ID250397, the bar starts to form rather late, around $t = 10$ Gyr, and the non-zero
signal in $A_3$ and $A_5$ (third and fifth panel) also starts to be visible around this time, while $A_1$ (first panel)
is then consistent with the noise. On the other hand, for ID511905, the bar is already forming at $t = 7$ Gyr, while any
asymmetry, in the form of non-zero $A_3$ and $A_5,$ starts to be visible later, around $t = 10$ Gyr. Again, very little
asymmetry in terms of $A_1$ is seen in the later stages.

In addition to this detailed study of the shape of the bars in terms of axis ratios and Fourier modes, we
also looked at their kinematics in order to determine if it is in any respect different in comparison with normal,
symmetric bars. It turns out that the velocity distribution in the lopsided bars in terms of mean velocity and
dispersion is very similar to that of symmetric bars; namely, there is a strong rotation signal in the edge-on and end-on view
and no significant rotation in the face-on projection \citep{Athanassoula2002, Lokas2020}. The maps of velocity
dispersion are also very much the same, with one strong maximum in the end-on view and two maxima in the edge-on and
face-on projection.

It is interesting to consider what might have caused the asymmetry of the bars. An obvious candidate is certainly
some kind of interaction with neighboring structures. In order to verify this hypothesis, we studied the
interactions of the selected sample of lopsided bars with other objects. In each case, we found mergers and flybys
of smaller and bigger structures, however, as a result of the nature of cosmological simulations, there are typically a
few interactions ongoing at the same time. Moreover, it takes time for the galaxy to respond dynamically to a given
interaction. As discussed above, the selected galaxies have very different evolutionary histories. Some are cluster
members that have therefore experienced encounters with other massive galaxies, albeit at high velocities. Others evolved in
relative isolation, experiencing only minor mergers. All these factors make it extremely difficult to identify a clear
pattern that would point toward a specific configuration.

Another scenario for the formation of lopsided bars is suggested by the evolution of the odd modes shown in
Fig.~\ref{oddmodestime}. As we have seen, in some simulation outputs, usually up to about $t = 7$ Gyr, all galaxies
show strong peaks in the $m = 1$ mode. We verified that in simulation outputs corresponding to the last peak
occurring in each galaxy, their stellar disks are indeed lopsided or off-center. The asymmetry is most probably caused by
off-center star formation in the asymmetrically distributed gas that is abundantly present in these galaxies at these
times. It is possible that these departures from symmetry affected the subsequent evolution of these galaxies, in
particular, the formation of the bar. In the next section, we consider these two scenarios using controlled simulations.

A further possibility is that lopsidedness develops in bars spontaneously in favorable conditions. Previous theoretical
works have demonstrated that asymmetric orbits of stars in bars can occur in strong-enough bars or with specific mass
distributions. In particular, \citet{Martinet1984} and \citet{Pfenniger1984}, in their studies of the orbital structure
of bars, identified asymmetric orbits as bifurcations of the standard x1 family that are believed to provide the main building
blocks of bars. It remains to be discovered under what conditions such orbits can dominate the structure of the
bar. The occurrence of these conditions must, however, be rare because most of the bars seen in the Universe and in the
simulations are indeed symmetric.

\section{Controlled simulations of the formation of lopsided bars}

\subsection{MW-like galaxy interacting with a satellite}

\begin{figure*}
\centering
\includegraphics[width=5.7cm]{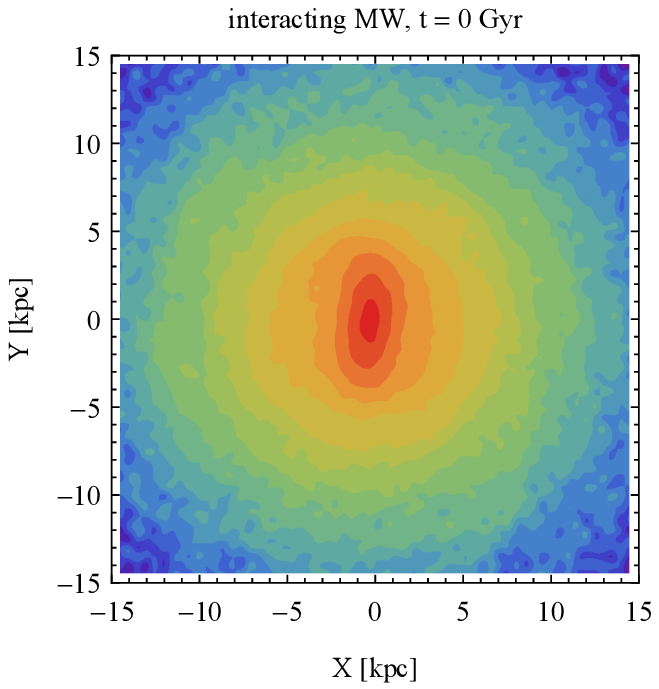}
\includegraphics[width=5.7cm]{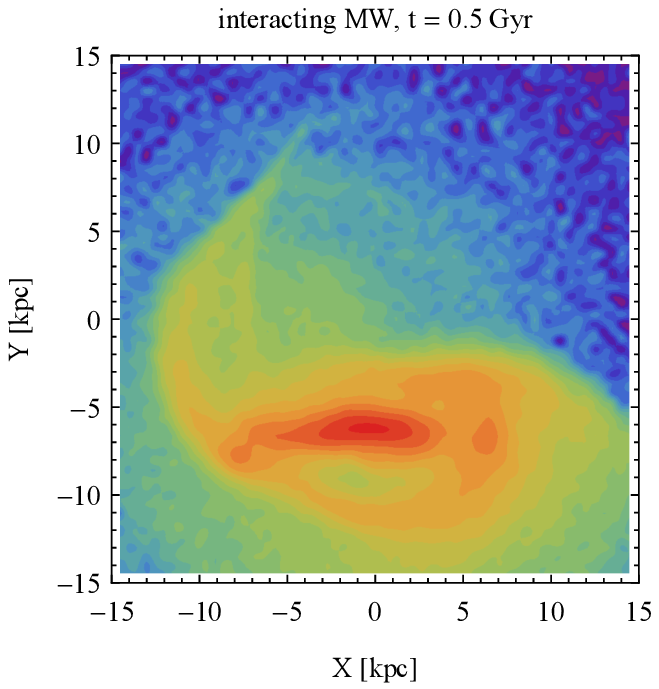}
\includegraphics[width=5.7cm]{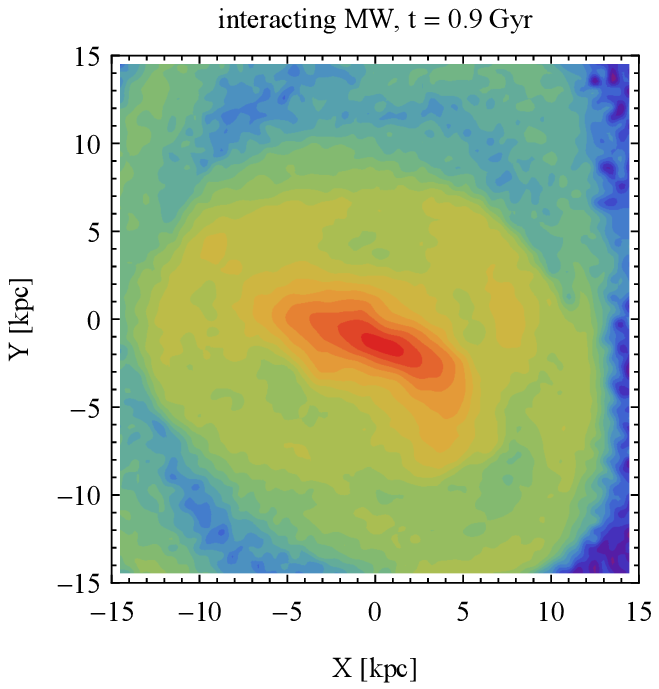} \\
\vspace{0.3cm}
\includegraphics[width=5.7cm]{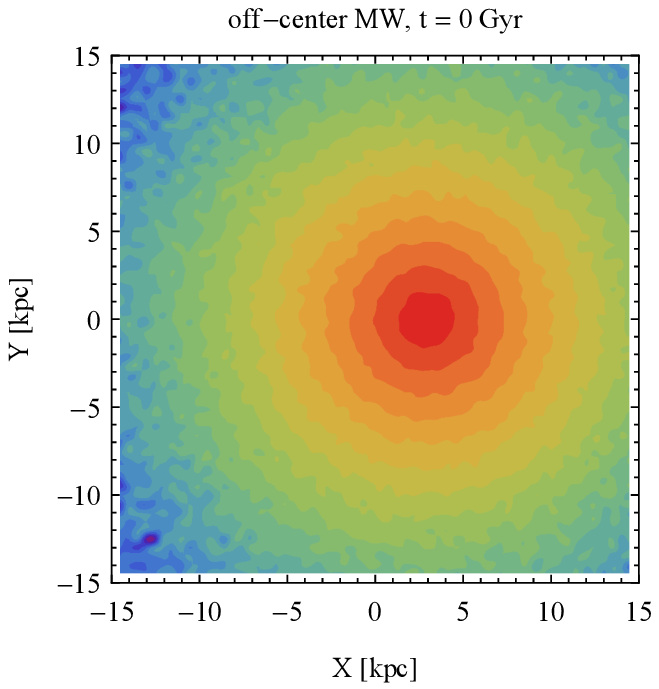}
\includegraphics[width=5.7cm]{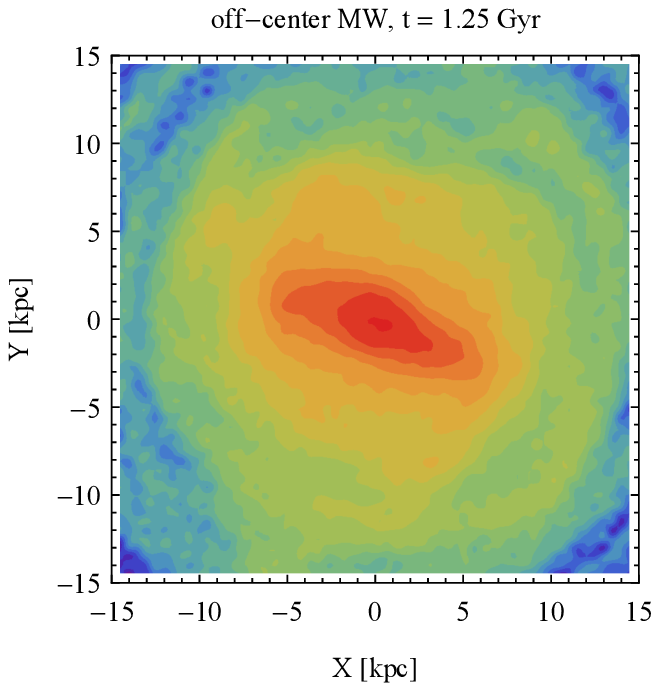}
\includegraphics[width=5.7cm]{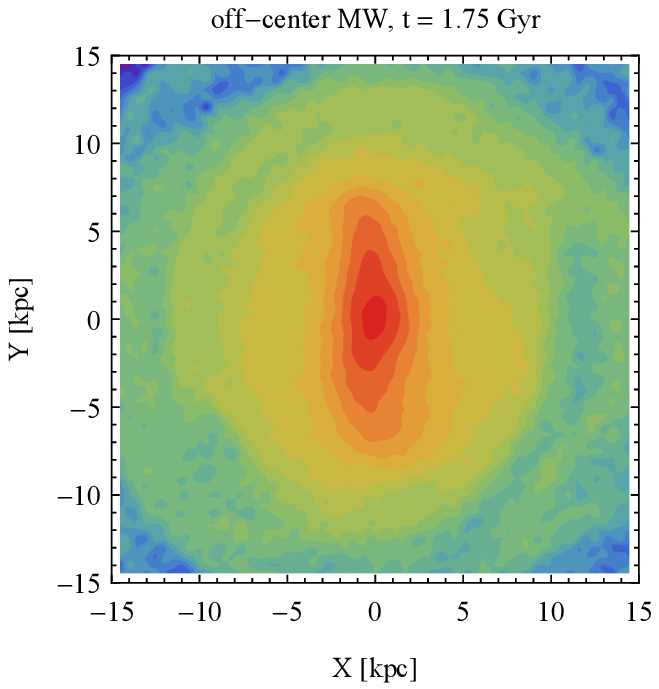}
\caption{Surface density distributions in the face-on view of the stellar components of the MW-like galaxies in
the controlled simulations. The coordinate system is that of the simulation box, with the galaxy disk always in the
$XY$ plane. In all images, the galaxy is rotating counter-clockwise. The three upper panels show three stages of the
evolution for the MW interacting with a satellite and the three lower ones three stages of the simulation of the MW
with an off-center disk. The initial configurations are presented in the left-column plots. The surface density,
$\Sigma,$ is normalized to the central maximum value in each case and the contours are equally spaced in $\log
\Sigma$.}
\label{surdenMW}
\end{figure*}

Trying to identify the mechanism responsible for the formation of a lopsided bar, we performed multiple controlled
simulations of a MW-like galaxy interacting with a satellite of different mass and orbit. In most cases, the
effect of such interactions on the forming or pre-existing bar is symmetric because of the approximately symmetric
nature of the tidal force acting on the host. Any distortions that occur in such interactions are very short-lived and
therefore do not explain the long-term asymmetries seen in the lopsided bars identified in IllustrisTNG.

The simulation that finally ended up producing an asymmetry, which we report below, was inspired to some extent by the recent
results of \citet{Collier2019} and \citet{Collier2021} who considered the effect of a counter-rotating halo on the bar
formation in the disk. They found that such a halo is much less efficient in transferring the angular
momentum from the disk than a corotating or isotropic one and, thus, the bar is forming more slowly. We  applied this
mechanism when constructing a controlled simulation of interacting galaxies and the formation of a lopsided bar, as
follows.

For the purpose of the simulation, we used the $N$-body model of the MW from \citet{Lokas2019}, which was used to study buckling instability in that work. The galaxy is made of two components: a spherical dark matter halo and an
exponential disk. The dark matter halo has an Navarro-Frenk-White \citep{Navarro1997} profile with a virial mass of
$M_{\rm H} = 10^{12}$ M$_{\odot}$ and concentration of $c=25,$ while the exponential disk had a mass of $M_{\rm D} = 4.5
\times 10^{10}$ M$_{\odot}$, scale-length of $R_{\rm D} = 3$ kpc, and thickness of $z_{\rm D} = 0.42$ kpc, with the central
value of the radial velocity dispersion of $\sigma_{R,0}=120$ km s$^{-1}$. The minimum value of the Toomre parameter
for this model at $2.5 R_{\rm D}$ is $Q=1.73,$ so the model was shown to be mildly unstable and forming a bar slowly
over the first few Gyr when evolving in isolation.

\begin{figure*}
\centering
\includegraphics[width=7cm]{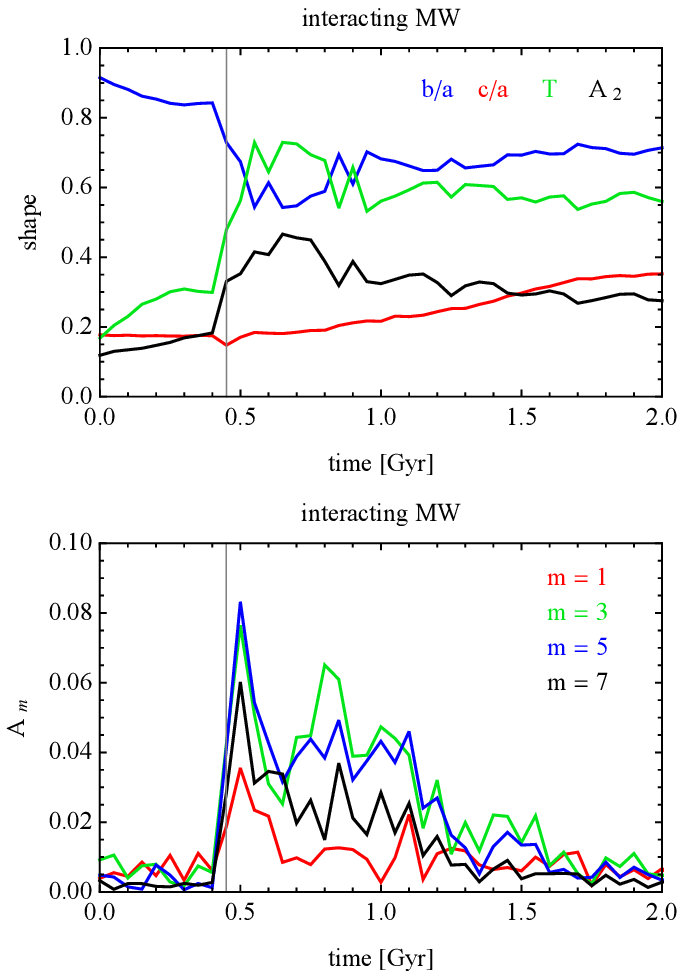}
\includegraphics[width=7cm]{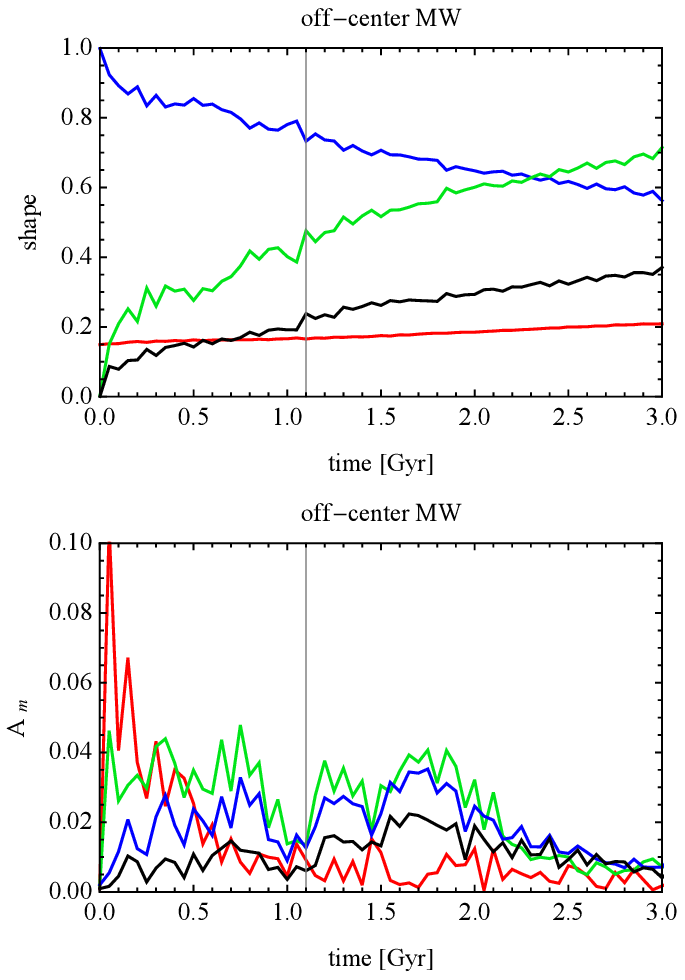}
\caption{Results of controlled simulations of the formation of lopsided bars in MW-like galaxies. Upper panels:
Evolution of the shape of the interacting MW (left) and the MW with an off-center disk (right). The blue and red lines
plot the intermediate to longest $b/a$ and the shortest to longest $c/a$ axis ratios, respectively. The green and black
lines show the evolution of the triaxiality parameter $T$ and the bar mode $A_2$. Lower panels: Odd Fourier modes of the
stellar component of the interacting MW (left) and the MW with an off-center disk (right) as a function of time. The
red, green, blue, and black lines show the results for $m = 1$, 3, 5, and 7, respectively. In all panels, the vertical
gray line indicates the time of bar formation.}
\label{shapeam}
\end{figure*}

The companion was assumed to contain only dark matter and exhibited a steep Navarro-Frenk-White profile with a an
exponential cut-off at 10 kpc and the total mass of $M_{\rm C} = 4.5 \times 10^{10}$ M$_{\odot}$, thus equal to the
disk mass of the host galaxy. In the inner parts, the density of the companion was comparable to the dark matter density
of the host. The $N$-body realizations of both galaxies were created with the procedures described in
\citet{Widrow2005} and \citet{Widrow2008} with each component of the host and the companion containing $10^6$
particles, respectively. The evolution of the interacting galaxies was followed with the GIZMO code
\citep{Hopkins2015}, an extension of the widely used GADGET-2 \citep{Springel2001, Springel2005}. The adopted
softening scales were $\epsilon_{\rm D} = 0.03$ kpc for the disk and the satellite, and $\epsilon_{\rm H} = 0.06$ kpc
for the halo of the galaxy, respectively.

For the initial conditions of this simulation we selected an output of the evolution of the MW-like galaxy in
isolation corresponding to $t = 2.25$ Gyr from the start, when the bar is already forming. The disk of the galaxy was
in the $XY$ plane of the simulation box, centered at $(X, Y, Z) = (0, 0, 0)$ kpc and was rotating so that its
angular momentum was along the positive $Z$ axis. This initial configuration is shown in the face-on
projection in the upper left panel of Fig.~\ref{surdenMW}. The companion was placed at $(X, Y, Z) = (0, -100, 0)$ kpc
and assigned the velocity $(V_X, V_Y, V_Z) = (0, 100, 0)$ km s$^{-1}$. It was therefore on the radial orbit toward the
galaxy, in the plane of the disk, that is moving upward along the $X=0$ axis in the upper left panel of
Fig.~\ref{surdenMW}. The evolution was followed for 2 Gyr, with outputs saved every 0.05 Gyr. With such initial
conditions, the collision, in terms of the first passage of the satellite through the center of the MW disk, takes place
after 0.5 Gyr. The upper middle panel of Fig.~\ref{surdenMW} shows the distorted bar just after the first
encounter with the satellite. We also note that the center of the bar is shifted toward negative $Y$ as a result of
the attraction from the satellite. Since the companion is bound to the galaxy, it soon turns around and oscillates
around its center, producing a system of approximately concentric shells that are expected to form in such collisions
\citep{Ebrova2012}.

The initial conditions were chosen so that at the time of the collision the bar forming in the MW is
perpendicular to the trajectory of the perturber. In such a configuration, at the time of the first passage, the satellite
acts differently on each side of the bar. On one side, the additional material contributed by the satellite
is moving in the same direction as the bar, thus acting as a co-rotating (prograde) halo and speeding up the formation
of the bar. On the other side, it is moving in the opposite direction to the bar, thus acting as a
counter-rotating (retrograde) halo and weakening the bar. The effect of the passage of the companion is thus different
on each side, resulting in the formation of a lopsided bar. An example of such a bar with significant
asymmetry, at $t = 0.9$ Gyr from the start of the simulation, is shown in the upper right panel of
Fig.~\ref{surdenMW}.

\begin{figure}[ht!]
\centering
\hspace{0.4cm}
\includegraphics[width=3.5cm]{legend24.eps}
\vspace{0.1cm}
\includegraphics[width=8.9cm]{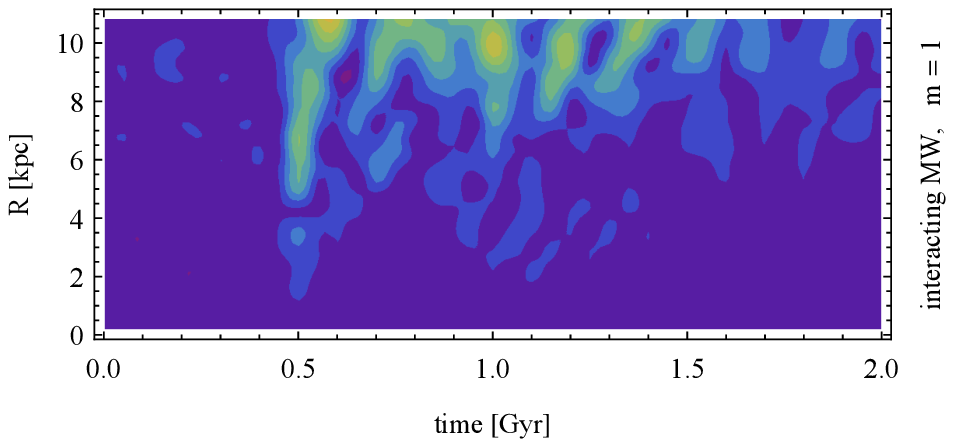}
\vspace{0.1cm}
\includegraphics[width=8.9cm]{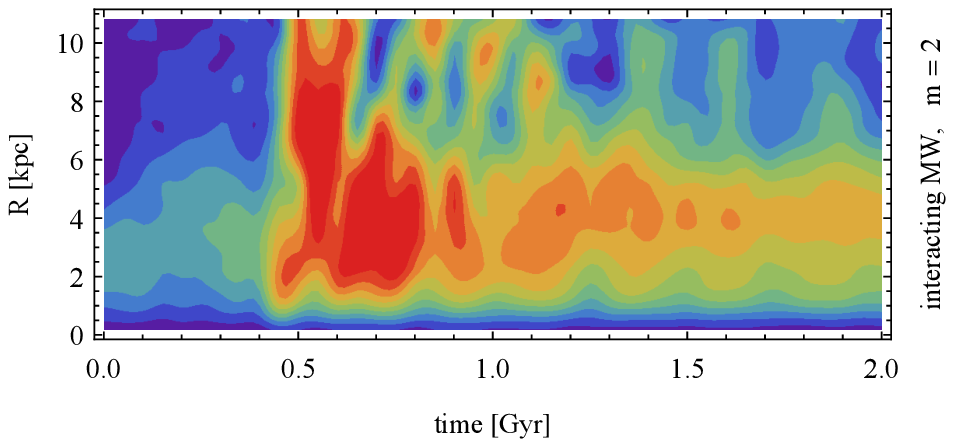}
\vspace{0.1cm}
\includegraphics[width=8.9cm]{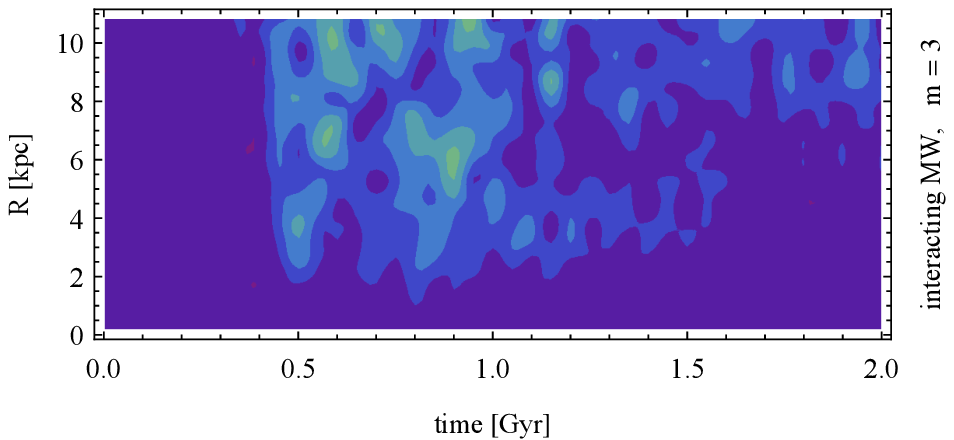}
\vspace{0.1cm}
\includegraphics[width=8.9cm]{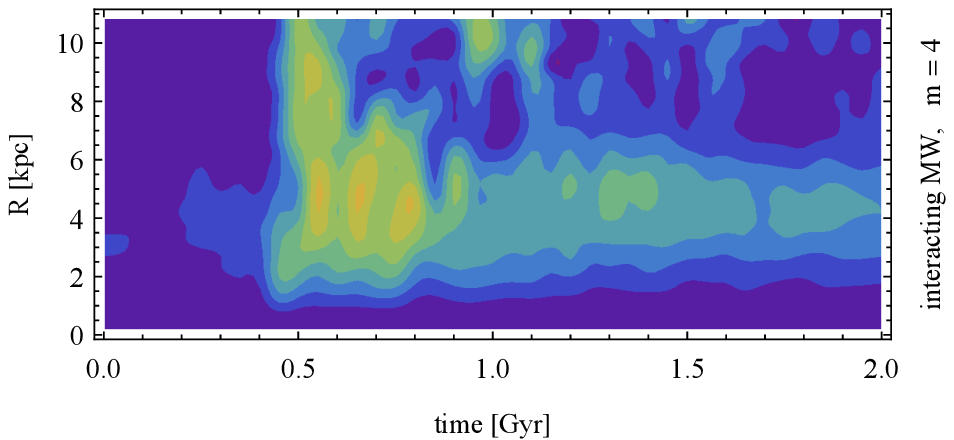}
\vspace{0.1cm}
\includegraphics[width=8.9cm]{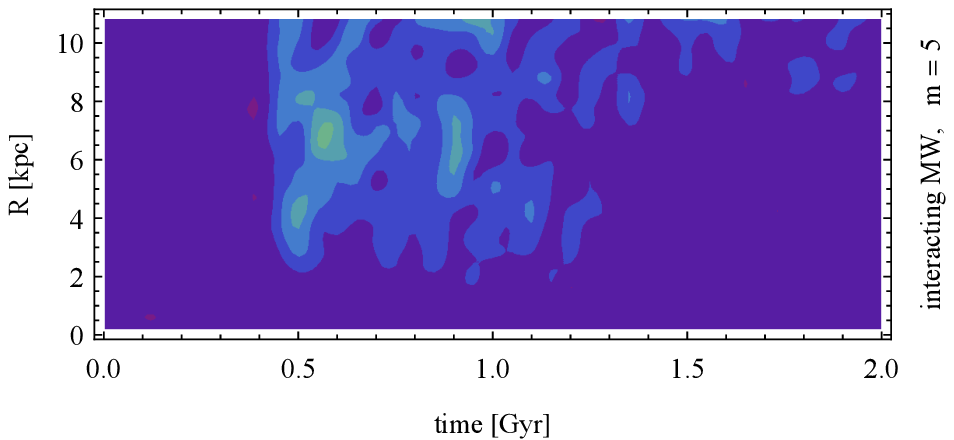}
\caption{Evolution of the profiles $A_m (R)$ over time for the controlled simulation of the
interacting MW. The five panels from top to bottom show the results for $m = 1$ to 5.}
\label{a2modestimeMW}
\end{figure}

\begin{figure}[ht!]
\centering
\hspace{0.4cm}
\includegraphics[width=3.5cm]{legend24.eps}
\vspace{0.1cm}
\includegraphics[width=8.9cm]{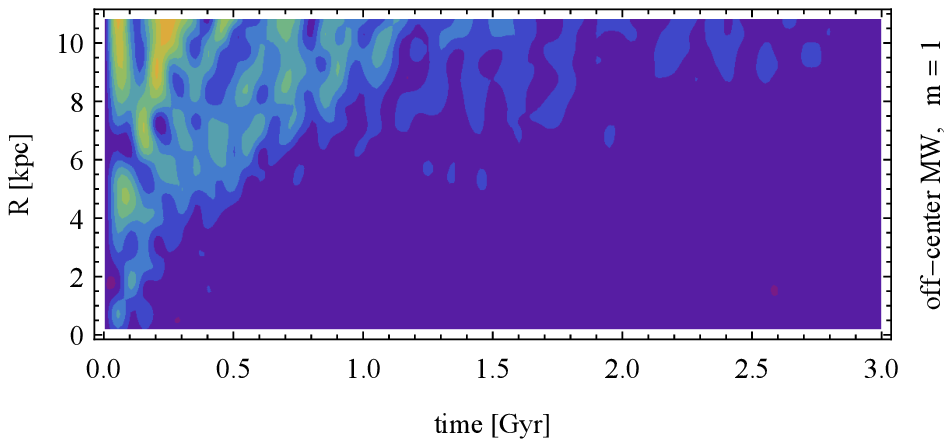}
\vspace{0.1cm}
\includegraphics[width=8.9cm]{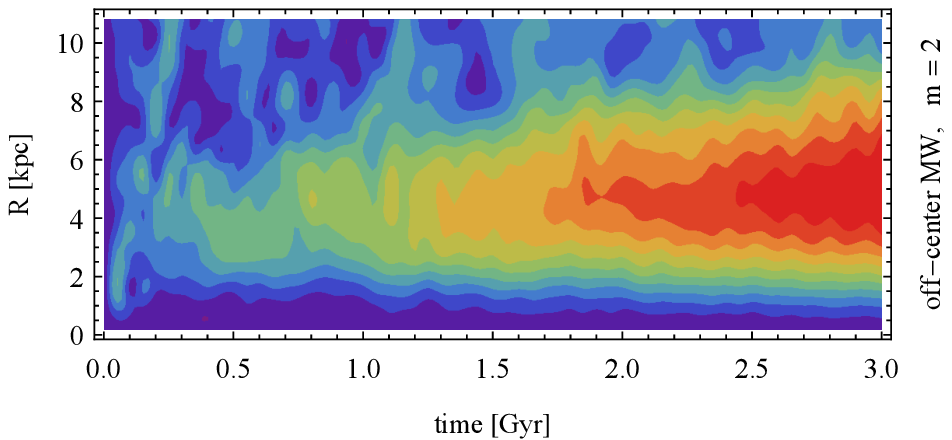}
\vspace{0.1cm}
\includegraphics[width=8.9cm]{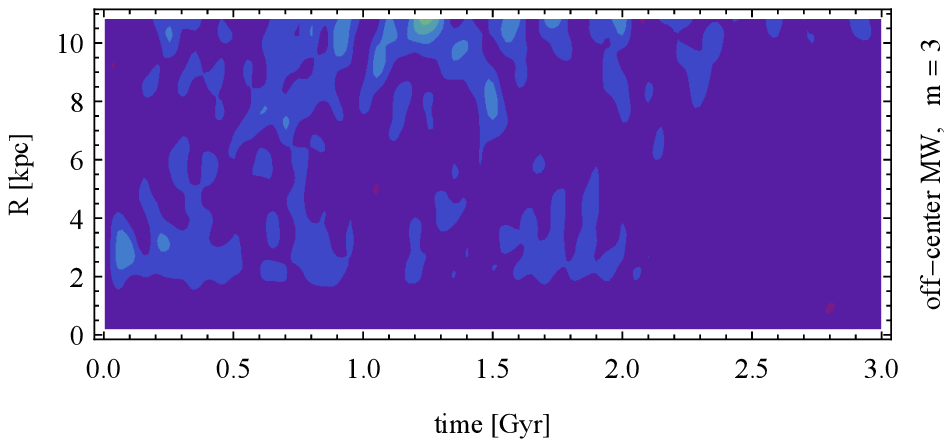}
\vspace{0.1cm}
\includegraphics[width=8.9cm]{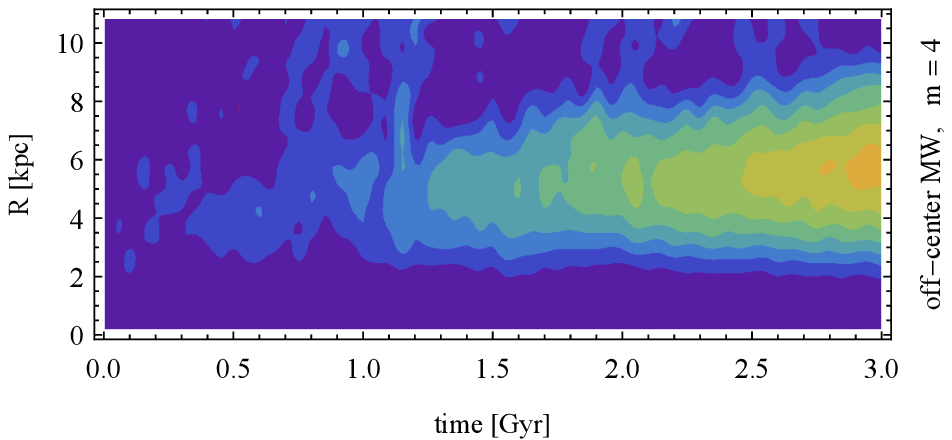}
\vspace{0.1cm}
\includegraphics[width=8.9cm]{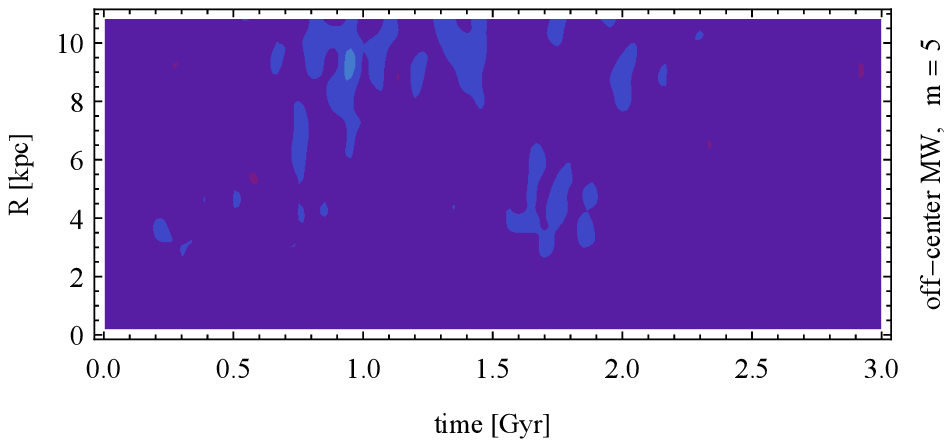}
\caption{Evolution of the profiles $A_m (R)$ over time for the controlled simulation of the MW with the
off-center disk. The five panels from top to bottom show the results for $m = 1$ to 5.}
\label{a2modestimeMWoff}
\end{figure}

The upper left panel of Fig.~\ref{shapeam} shows the evolution of different measures of shape of the interacting galaxy
in a way that is similar to Fig.~\ref{shape} for the IllustrisTNG galaxies. The axis ratios $b/a$, $c/a$, the triaxiality $T,$
and the bar mode $A_2$ were calculated using MW stars within $2 R_{\rm D} = 6$ kpc with the center of the
stellar mass determined iteratively in spheres of decreasing size down to the radius of 0.5 kpc. The vertical gray
line marks the formation of the bar, namely, the time when $A_2$ becomes larger than 0.2. This is also the time when the
bar starts to grow rapidly as a result of the interaction with the satellite. It then regains equilibrium with $A_2$
remaining at the level of about 0.3, and later around $t = 1.5$ Gyr, it experiences a weak buckling episode.

The lower left panel of Fig.~\ref{shapeam} shows the evolution of the odd Fourier modes $A_m,$ with $m = 1$, 3, 5, and 7
in a way that is similar to Fig.~\ref{oddmodestime} for IllustrisTNG galaxies. We see that the modes start to grow rapidly at
the time of interaction and preserve their enhanced values for about 1 Gyr. Later they decrease and reach the
pre-interaction level at about $t = 1.5$ Gyr, that is at the time of buckling. The hierarchy of the modes is similar to
that of the IllustrisTNG galaxies at the later stages, namely, the $A_1$ mode is subdominant, while $A_3$ and
$A_5$ have the highest values.

A fuller picture of the bar properties can be obtained by viewing the color-coded images of the dependence of $A_m$
modes on both radius and time. These are shown in Fig.~\ref{a2modestimeMW}, in a set of plots similar to those of
Figs.~\ref{a2modestime1}-\ref{a2modestime2} for two IllustrisTNG galaxies. Here, the bin size in radius is
again $\Delta R = 0.5$ kpc and the subsequent snapshots differ by $\Delta t = 0.05$ Gyr. As before, the color map of
the evolution of $A_2$ can be used to estimate the bar length at different times by finding the radius where the $A_2$
value drops to half its maximum. As expected, the length of the bar is very much increased during the interaction as
the bar is stretched, but then it settles to a value of about 6 kpc. The asymmetry as measured by $A_1$ dominates in
the outskirts of the disk, outside the bar, while $A_3$ and $A_5$ have the highest values between $R = 2$ and $R = 8$
kpc.

We compared the evolution of the odd modes in the interacting case to the case when the MW evolves in isolation by
measuring $A_m$ in the same way. We found that in this case, $A_3$ and $A_5$ increase slightly for a short period of
time, reaching values of about 0.02. This happens just before buckling, after which the odd modes decrease to the noise
level again. This short occurrence of the weak asymmetry may be related to the appearance of orbits with a loop on one
end, as shown for the one example orbit plotted in the upper left panel of Fig. 9 in \citet{Lokas2019}. After buckling,
the orbital structure is completely rebuilt and asymmetric orbits seem to disappear. The same seems to occur in the
interacting MW model discussed here since the asymmetry also disappears after buckling, although in this case, the
buckling is much weaker. A few examples of orbits contributing to the lopsided shape of the bar are presented
in the appendix.

\subsection{MW-like galaxy with an off-center disk}

Trying to assess the role of interactions in creating off-center bars, \citet{Kruk2017} searched for neighbors around
the galaxies with off-center bars in their sample and concluded that, surprisingly, a significant fraction of them were
isolated. These authors noted that this phenomenon may be caused by the misalignment between the disk and the halo, as
discussed by \citet{Levine1998}, who performed simulations of the evolution of a galaxy with a disk shifted with
respect to the halo. As is visible in their Fig. 1, such a configuration produces not only an off-center, but also a
lopsided bar. Since we found signatures of such off-center disks in our galaxies with lopsided bars from IllustrisTNG,
here we take a closer look at how such a configuration may affect the properties of the forming bar.

In order to verify whether this is a viable scenario for the formation of a lopsided bar, we performed
simulations of our MW model with the disk displaced with respect to the halo center. For the purpose of these
simulations, we used the same MW model as described in the previous subsection, initially containing  an exponential disk
and a dark matter halo. The disk was shifted by one disk scale-length $R_{\rm D} = 3$ kpc in the direction of positive
$X$ axis of the simulation box, as shown in the lower left panel of Fig.~\ref{surdenMW}. We tried smaller and
larger shifts as well and found that shifts of $R_{\rm D}/3 = 1$ kpc and $2 R_{\rm D}/3 = 2$ kpc lead to very little
asymmetry in the formed bars while a larger shift of $2 R_{\rm D} = 6$ kpc leads to the formation of an elongated
structure but with a large bulge-like component in the center that does not especially resemble a bar. Therefore, in this work, we
describe only the simulation with the shift of one disk scale-length, evolved for 5 Gyr. This configuration indeed led
to the formation of a lopsided bar, as confirmed by the face-on images of the stellar component shown in the lower
middle and lower right panels of Fig.~\ref{surdenMW}.

In the upper right panel of Fig.~\ref{shapeam} we plot the evolution of the measures of shape of this galaxy:
the axis ratios, $b/a$, $c/a$, the triaxiality, $T,$ and the bar mode, $A_2,$ calculated in the same way as for the
interacting model, but up to 3 Gyr. The vertical gray line marks again the time of the formation of the bar, when $A_2$
grows above 0.2. We see that except for the very beginning, the evolution is now less violent, with the $b/a$ ratio
decreasing and the remaining parameters slowly increasing.

The lower right panel of Fig.~\ref{shapeam} shows the evolution of the odd Fourier modes $A_m,$ with $m = 1$, 3,
5, and 7. We see that in contrast to the interacting model, the $m = 1$ mode grows strongly at the beginning, signifying
the temporary presence of a lopsided disk as it falls into the potential well of the dark halo. Once the galaxy settles
into equilibrium and the bar starts to grow, this mode becomes subdominant with respect to the other odd modes. The
values of $A_3$ and $A_5$ retain significant non-zero values after the bar formation for about 1 Gyr, which means that
the bar is indeed lopsided. After about 2 Gyr of evolution, they start to decrease and the bar becomes more symmetric.
Contrary to the case of the interacting galaxy, this behavior is not associated with buckling because in
this case the buckling occurs as late as around 4 Gyr, but it does further decrease the odd modes.

In Fig.~\ref{a2modestimeMWoff}, we show the color-coded images of the dependence of $A_m$ modes on
both radius and time. The bin sizes in radius and time are the same as for the interacting model, $\Delta R = 0.5$ kpc
and $\Delta t = 0.05$ Gyr. The evolution of $A_2$ and $A_4$ (the second and fourth panels) confirms the steady
growth of the bar and the map for $A_2$ can be used to estimate the bar length at different times by finding the radius
where the $A_2$ value drops to half its maximum. Contrary to the case of the interacting MW, the length of the bar
increases steadily and at $t = 1.75$ Gyr, it is equal to about 8 kpc, which is in agreement with the visual impression from the
lower right panel of Fig.~\ref{surdenMW}. The first panel of Fig.~\ref{a2modestimeMWoff} shows the map for $A_1$
and confirms the high initial values of the $m = 1$ mode due to the presence of the lopsided disk. Later on, only the $A_3$
and $A_5$ odd modes show non-zero values in the bar region, between the radii $R = 2$ kpc and $R = 6$ kpc.

\section{Discussion}

In this work, we present a few convincing examples of lopsided bars formed in simulations of galaxy evolution in
the cosmological context. The bars are lopsided in the sense that they possess a significant degree of asymmetry in their
face-on images. These objects differ from the well-known class of Magellanic-type galaxies that, instead, have off-center bars that are embedded in asymmetric disks. We show here that the asymmetry of the bars is a long-lived feature, rather than a transient
one, and can be sustained for a few Gyr. The asymmetry can grow along with the bar or it can occur later in its
evolution.

An obvious candidate for the mechanism inducing the asymmetry of the bars is some sort of interaction with another
galaxy. By tracing the interactions of the lopsided bars identified in IllustrisTNG, we identified a number of flybys and
mergers for each of them, but failed to pinpoint a configuration that they all have in common that could lead to the
asymmetry. Using a controlled simulation of a MW-like galaxy interacting with a massive companion, we propose a
particular mechanism that leads to the formation of lopsidedness in the bar. In our simulation, the satellite is placed on a radial orbit in the
plane of the disk so that its trajectory is perpendicular to the already forming bar. Although the satellite is as
massive as the MW disk, the bar survives the interaction --  and it is even enhanced. At the same time, because of the
asymmetry in the effects of the satellite on both sides of the bar, it becomes lopsided. The asymmetry is retained until
the bar buckles, which suggests that during buckling, the orbital structure is rebuilt so that it becomes again
symmetric in the face-on view.

We also considered an alternative scenario for the formation of a lopsided bar inspired by the study of
\citet{Levine1998}, aimed at reproducing the off-center bars. In this model, the galaxy disk is initially displaced with
respect to the dark matter halo. The bar forming in such a configuration also shows some degree of asymmetry. The
feature persists for a significant amount of time but is weaker and disappears even before buckling. However, the exact
amount of asymmetry may depend on many parameters that were not varied here.

Comparing the evolution of the $A_m$ modes in the two scenarios shown in Figs.~\ref{shapeam},
\ref{a2modestimeMW}, and \ref{a2modestimeMWoff} to the analogous results for the IllustrisTNG galaxies in
Figs.~\ref{shape}, \ref{oddmodestime}, \ref{a2modestime1}, and \ref{a2modestime2}, we see that the time dependence of the
modes in IllustrisTNG bars is more similar to the scenario with the off-center disk. The even modes grow more gradually
while the odd modes show an early peak in the $m = 1$ mode followed by an increase in the higher odd modes. The
situation involving the scenario with the interacting MW is different: there is a sharp increase in the bar mode, but the $m =
1$ mode is never very strong.

We may therefore conclude that the scenario with the off-center disk is more probable for the formation of lopsided
bars. This would mean that the bars inherit their lopsidedness from the disks. The caveat here is that at the time of
the occurrence of the last $m = 1$ peak in IllustrisTNG galaxies, they are not yet forming bars (except for
ID177725) and the $m = 1$ peaks are very short-lived, so it is hard to imagine a strong causal relation between the
lopsided disks and lopsided bars. This problem is particularly evident in the case of ID511905, where the bar forms and
its asymmetry starts to be visible almost 5 Gyr after the last peak in $A_1$.

Although the LMC is considered to be a prototype of Magellanic-type galaxies with off-center bars, its deprojected
face-on image \citep{Marel2001} suggests that its bar is not only off-center with respect to the disk, but also
lopsided. A more recent study by \citet{Jacyszyn2016}, based on cepheids from the OGLE project, has found that the
LMC bar is probably not strongly off-center but still asymmetric. LMC is thus the first known, convincing example of a
lopsided bar. It is possible that the two phenomena are, in fact, related, although the existing simulations aiming to
reproduce the shape of LMC fail to form a lopsided bar, but they do succeed in having it off-center with respect to the disk.
Another example of a galaxy with a lopsided bar is the dwarf irregular DDO 168, as discussed by \citet{Patra2019}. It
remains to be seen how many of the Magellanic-type galaxies turn out to have their bars not only off-center, but also
lopsided. At present, the quality of their available images appears to be insufficient with regard  to discerning the details of the
bar structure.

We may certainly wonder whether the bar of the MW might be lopsided in the sense discussed in this work. Although the image of
the MW bar revealed by APOGEE DR16 and {\em Gaia} DR2 is asymmetric \citep{Queiroz2020}, this is mainly due to the
different extinction on both sides of the bar. We also note that the view of the MW bar that is available to us is edge-on,
while the asymmetry discussed here is mainly visible in face-on images. In addition, the MW bar is known to possess a
clear boxy/peanut shape \citep{Dwek1995, Ness2016} indicating that it must have undergone a buckling episode in the
past; as discussed above, this would probably have erased any asymmetry present in the bar in the earlier stages.

\begin{acknowledgements}
I am grateful to the anonymous referee for useful comments and to the IllustrisTNG team for making their
simulations publicly available.
\end{acknowledgements}

\begin{appendix}
\section{Orbits of stars in the lopsided bar}

\begin{figure}
\centering
\includegraphics[width=8.9cm]{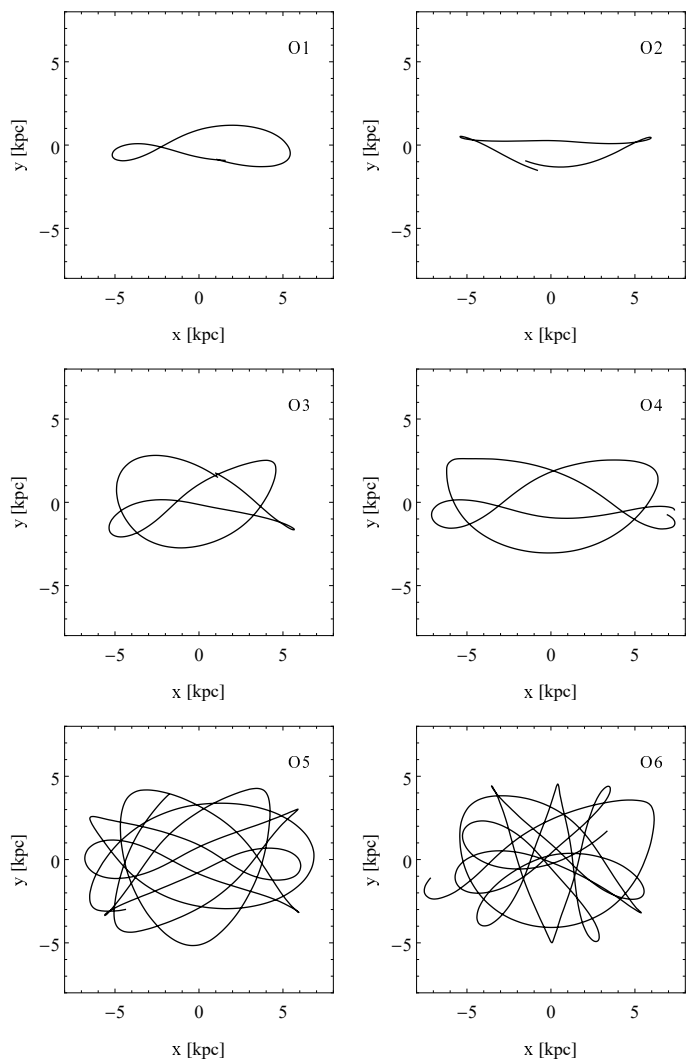}
\caption{Examples of stellar orbits from the simulation of the interacting MW contributing to the lopsided shape of
the bar.}
\label{orbits}
\end{figure}

It is interesting to consider what orbits could be responsible for the asymmetry of the lopsided bar. Although
the study of the detailed orbital structure is beyond the scope of the present paper, we provide here a few examples for
the case of the bar forming in our simulation of the MW-like galaxy interacting with a satellite described in
Section 3.1. To trace the orbits, we reran the simulation between $t = 0.5$ and $t = 1.5$ Gyr, when the asymmetry is
present, saving outputs every 0.001 Gyr.

A few typical examples of orbits that may be responsible for the asymmetry are presented in Fig.~\ref{orbits}. The upper
four panels show orbits for a fraction of the resimulated 1 Gyr to display  their shapes clearly. The two lower panels
show more complicated orbits for the maximum time of 1 Gyr and we see that they have not yet completed a full period.
The upper two orbits are two examples familiar from the studies of \citet{Martinet1984} and \citet{Pfenniger1984}. In
the latter work, the orbits O1 and O2 in the upper row of Fig.~\ref{orbits} were called $P$ and $C$, respectively, and
classified as bifurcations of the standard x1 family. From the two lower rows of Fig.~\ref{orbits}, we see that there
are many more types of orbits and of much more complicated shapes contributing to the formation of a lopsided bar.

\end{appendix}


\begin{thebibliography}{}

\bibitem[{Athanassoula \& Misiriotis}(2002)]{Athanassoula2002} Athanassoula, E., \& Misiriotis, A. 2002,
        MNRAS, 330, 35
\bibitem[{Athanassoula et al.}(2013)]{Athanassoula2013} Athanassoula, E., Machado, R. E. G., \& Rodionov, S. A. 2013,
        MNRAS, 429, 1949
\bibitem[{Bekki}(2009)]{Bekki2009} Bekki, K. 2009, MNRAS, 393, L60
\bibitem[{Besla et al.}(2012)]{Besla2012} Besla, G., Kallivayalil, N., Hernquist, L., et al. 2012, MNRAS, 421, 2109
\bibitem[{Bournaud et al.}(2005)]{Bournaud2005} Bournaud, F., Combes, F., Jog, C. J., \& Puerari, I. 2005,
        A\&A, 438, 507
\bibitem[{Buta et al.}(2015)]{Buta2015} Buta, R. J., Sheth, K., Athanassoula, E., et al. 2015, ApJS, 217, 32
\bibitem[{Collier \& Madigan}(2021)]{Collier2021} Collier, A., \& Madigan, A.-M., 2021, ApJ, 915, 23
\bibitem[{Collier et al.}(2019)]{Collier2019} Collier, A., Shlosman, I., \& Heller, C. 2019, MNRAS, 489, 3102
\bibitem[{Debattista et al.}(2006)]{Debattista2006} Debattista, V. P., Mayer, L., Carollo, C. M., et al.
        2006, ApJ, 645, 209
\bibitem[{D\'{i}az-Garc\'{i}a et al.}(2016)]{Diaz2016} D\'{i}az-Garc\'{i}a, S., Salo, H., Laurikainen, E., \&
        Herrera-Endoqui, M. 2016, A\&A, 587, A160
\bibitem[{Dwek et al.}(1995)]{Dwek1995} Dwek, E., Arendt, R. G., Hauser, M. G., et al. 1995, ApJ, 445, 716
\bibitem[{Ebrov\'{a} et al.}(2012)]{Ebrova2012} Ebrov\'{a}, I., J\'{i}lkov\'{a}, L., Jungwiert, B., et al.
        2012, A\&A, 545, A33
\bibitem[{Genel et al.}(2015)]{Genel2015} Genel, S., Fall, S. M., Hernquist, L., et al. 2015, ApJ, 804, L40
\bibitem[{Genel et al.}(2018)]{Genel2018} Genel, S., Nelson, D., Pillepich, A., et al. 2018, MNRAS, 474, 3976
\bibitem[{Gerin et al.}(1990)]{Gerin1990} Gerin, M., Combes, F., \& Athanassoula, E. 1990, A\&A, 230, 37
\bibitem[{Ghosh et al.}(2021)]{Ghosh2021} Ghosh, S., Saha, K., Jog, C. J., Combes, F., \& Di Matteo, P. 2021,
        arXiv:2105.05270
\bibitem[{Hohl}(1971)]{Hohl1971} Hohl, F. 1971, ApJ, 168, 343
\bibitem[{Hopkins}(2015)]{Hopkins2015} Hopkins, P. F. 2015, MNRAS, 450, 53
\bibitem[{Jacyszyn-Dobrzeniecka et al.}(2016)]{Jacyszyn2016} Jacyszyn-Dobrzeniecka, A. M., Skowron, D. M., Mr\'{o}z, P.,
        et al. 2016, AcA, 66, 149
\bibitem[{Jog \& Combes}(2009)]{Jog2009} Jog, C. J. , \& Combes, F. 2009, Phys. Rep., 471, 75
\bibitem[{Kruk et al.}(2017)]{Kruk2017} Kruk, S. J., Lintott, C. J., Simmons, B. D., et al. 2017, MNRAS, 469, 3363
\bibitem[{Levine \& Sparke}(1998)]{Levine1998} Levine, S. E., \& Sparke, L. S. 1998, ApJ, 496, L13
\bibitem[{{\L}okas}(2018)]{Lokas2018} {\L}okas, E. L. 2018, ApJ, 857, 6
\bibitem[{{\L}okas}(2019)]{Lokas2019} {\L}okas, E. L. 2019, A\&A, 629, A52
\bibitem[{{\L}okas}(2020)]{Lokas2020} {\L}okas, E. L. 2020, A\&A, 638, A133
\bibitem[{{\L}okas}(2021)]{Lokas2021} {\L}okas, E. L. 2021, A\&A, 647, A143
\bibitem[{{\L}okas et al.}(2014)]{Lokas2014} {\L}okas, E. L., Athanassoula, E., Debattista, V. P., et al. 2014,
        MNRAS, 445, 1339
\bibitem[{Mapelli et al.}(2008)]{Mapelli2008} Mapelli, M., Moore, B., \& Bland-Hawthorn, J. 2008, MNRAS, 388, 697
\bibitem[{Marinacci et al.}(2018)]{Marinacci2018} Marinacci, F., Vogelsberger, M., Pakmor, R., et al. 2018,
        MNRAS, 480, 5113
\bibitem[{Martinet}(1984)]{Martinet1984} Martinet, L. 1984, A\&A, 132, 381
\bibitem[{Melvin et al.}(2014)]{Melvin2014} Melvin, T., Masters, K., Lintott, C., et al. 2014, MNRAS, 438, 2882
\bibitem[{Miwa \& Noguchi}(1998)]{Miwa1998} Miwa, T., \& Noguchi, M. 1998, ApJ, 499, 149
\bibitem[{Naiman et al.}(2018)]{Naiman2018} Naiman, J. P., Pillepich, A., Springel, V., et al., 2018, MNRAS, 477, 1206
\bibitem[{Navarro et al.}(1997)]{Navarro1997} Navarro, J. F., Frenk, C. S., \& White, S. D. M. 1997, ApJ, 490, 493
\bibitem[{Nelson et al.}(2018)]{Nelson2018} Nelson, D., Pillepich, A., Springel, V., et al. 2018, MNRAS, 475, 624
\bibitem[{Nelson et al.}(2019)]{Nelson2019} Nelson, D., Springel, V., Pillepich, A., et al. 2019,
        Computational Astrophysics and Cosmology, 6, 2
\bibitem[{Ness \& Lang}(2016)]{Ness2016} Ness, M., \& Lang, D. 2016, AJ, 152, 14
\bibitem[{Noguchi}(1987)]{Noguchi1987} Noguchi, M. 1987, MNRAS, 228, 635
\bibitem[{Odewahn}(1994)]{Odewahn1994} Odewahn, S. C. 1994, AJ, 107, 1320
\bibitem[{Ostriker \& Peebles}(1973)]{Ostriker1973} Ostriker, J. P., \& Peebles, P. J. E. 1973, ApJ, 186, 467
\bibitem[{Pardy et al.}(2016)]{Pardy2016} Pardy, S. A., D'Onghia, E., Athanassoula, E., Wilcots, E. M., \& Sheth, K.
        2016, ApJ, 827, 149
\bibitem[{Patra \& Jog}(2019)]{Patra2019} Patra, N. N., \& Jog, C. J. 2019, MNRAS, 488, 4942
\bibitem[{Peschken \& {\L}okas}(2019)]{Peschken2019} Peschken, N., \& {\L}okas, E. L. 2019, MNRAS, 483, 2721
\bibitem[{Pfenniger}(1984)]{Pfenniger1984} Pfenniger, D. 1984, A\&A, 134, 373
\bibitem[{Pillepich et al.}(2018)]{Pillepich2018} Pillepich, A., Nelson, D., Hernquist, L., et al. 2018,
        MNRAS, 475, 648
\bibitem[{Queiroz et al.}(2020)]{Queiroz2020} Queiroz, A. B. A., Chiappini, C., Perez-Villegas, A., et al. 2020,
        submitted to A\&A, arXiv:2007.12915
\bibitem[{Rix \& Zaritsky}(1995)]{Rix1995} Rix, H.-W., \& Zaritsky, D. 1995, ApJ, 447, 82
\bibitem[{Rodriguez-Gomez et al.}(2019)]{Rodriguez2019} Rodriguez-Gomez, V.,  Snyder, G. F., Lotz, J. M., et al.
        2019, MNRAS, 483, 4140
\bibitem[{Rosas-Guevara et al.}(2020)]{Rosas2020} Rosas-Guevara, Y., Bonoli, S., Dotti, M., et al. 2020, MNRAS, 491,
        2547
\bibitem[{Sellwood}(1981)]{Sellwood1981} Sellwood, J. A. 1981, A\&A, 99, 362
\bibitem[{Sheth et al.}(2008)]{Sheth2008} Sheth, K., Elmegreen, D. M., Elmegreen, B. G., et al. 2008, ApJ, 675, 1141
\bibitem[{Skibba et al.}(2012)]{Skibba2012} Skibba, R. A., Masters, K. L., Nichol, R. C. 2012, MNRAS, 423, 1485
\bibitem[{Springel}(2005)]{Springel2005} Springel, V. 2005, MNRAS, 364, 1105
\bibitem[{Springel et al.}(2001)]{Springel2001} Springel, V., Yoshida, N., \& White, S. D. M. 2001,
        New Astronomy, 6, 79
\bibitem[{Springel et al.}(2018)]{Springel2018} Springel, V., Pakmor, R., Pillepich, A., et al. 2018, MNRAS, 475, 676
\bibitem[{van der Marel}(2001)]{Marel2001} van der Marel, R. P. 2001, AJ, 122, 1827
\bibitem[{Widrow \& Dubinski}(2005)]{Widrow2005} Widrow, L. M., \& Dubinski, J. 2005, ApJ, 631, 838
\bibitem[{Widrow et al.}(2008)]{Widrow2008} Widrow, L. M., Pym, B., \& Dubinski, J. 2008, ApJ, 679, 1239
\bibitem[{Zaritsky et al.}(2013)]{Zaritsky2013} Zaritsky, D., Salo, H., Laurikainen, E., et al. 2013, ApJ, 772, 135
\bibitem[{Zhao et al.}(2020)]{Zhao2020} Zhao, D., Du, M., Ho, L. C., Debattista, V. P., \& Shi, J. 2020,
        ApJ, 904, 170
\bibitem[{Zhou et al.}(2020)]{Zhou2020} Zhou, Z.-B., Zhu, W., Wang, Y., \& Feng, L.-L. 2020, ApJ, 895, 92


\end{thebibliography}
\end{document}